\documentclass[
  journal=pasa,
  manuscript=research-paper, 
  year=2020,
  volume=37,
]{cup-journal}

\usepackage{microtype,siunitx,booktabs}
\usepackage[final]{changes}

\title{MWAX: A New Correlator for the Murchison Widefield Array}

\author{I. S. Morrison}
\affiliation{International Centre for Radio Astronomy Research (ICRAR), Curtin University, Bentley, WA 6102, Australia}
\email[I. S. Morrison]{ian.morrison@curtin.edu.au}

\author{B. Crosse}
\affiliation{International Centre for Radio Astronomy Research (ICRAR), Curtin University, Bentley, WA 6102, Australia}

\author{G. Sleap}
\affiliation{International Centre for Radio Astronomy Research (ICRAR), Curtin University, Bentley, WA 6102, Australia}

\author{R. B. Wayth}
\affiliation{International Centre for Radio Astronomy Research (ICRAR), Curtin University, Bentley, WA 6102, Australia}

\author{A. Williams}
\affiliation{International Centre for Radio Astronomy Research (ICRAR), Curtin University, Bentley, WA 6102, Australia}

\author{M. Johnston-Hollitt}
\affiliation{Curtin Institute for Computation, Curtin University, GPO Box U1987, Perth WA 6845, Australia}
\alsoaffiliation{International Centre for Radio Astronomy Research (ICRAR), Curtin University, Bentley, WA 6102, Australia}

\author{J. Jones}
\affiliation{International Centre for Radio Astronomy Research (ICRAR), Curtin University, Bentley, WA 6102, Australia}

\author{S. J. Tingay}
\affiliation{International Centre for Radio Astronomy Research (ICRAR), Curtin University, Bentley, WA 6102, Australia}

\author{M. Walker}
\affiliation{International Centre for Radio Astronomy Research (ICRAR), Curtin University, Bentley, WA 6102, Australia}

\author{L. Williams}
\affiliation{International Centre for Radio Astronomy Research (ICRAR), Curtin University, Bentley, WA 6102, Australia}

\doi{10.1017/pasa.2020.32}

\received {8 July 2022}
\revised  {24 January 2023}
\accepted {22 February 2023}
\published{dd Mmm YYYY}


\begin{document}

\begin{abstract}

We describe the design, validation, and commissioning of a new correlator termed ``MWAX'' for the Murchison Widefield Array (MWA) low-frequency radio telescope.  MWAX replaces an earlier generation MWA correlator, extending correlation capabilities and providing greater flexibility, scalability, and maintainability.  MWAX is designed to exploit current and future Phase II/III upgrades to MWA infrastructure, most notably the simultaneous correlation of all 256 of the MWA's antenna tiles (and potentially more in future).  MWAX is a fully software-programmable correlator based around an ethernet multicast architecture.  At its core is a cluster of 24 high-performance GPU-enabled commercial-off-the-shelf compute servers that together process in real-time up to 24 coarse channels of 1.28 MHz bandwidth each.  The system is highly flexible and scalable in terms of the number of antenna tiles and number of coarse channels to be correlated, and it offers a wide range of frequency / time resolution combinations to users.  We conclude with a roadmap of future enhancements and extensions that we anticipate will be progressively rolled out over time.
\end{abstract}

\section{INTRODUCTION}

Interferometric array radio telescopes like the Murchison Wi-defield Array (MWA) can form detailed images of the sky through a complex process involving multiple stages of signal processing and data manipulation.
At the heart of the process is the ``correlator'' – a computational engine that calculates what are known as ``visibilities'', which are the cross-correlations between every pair of array elements \citep{1999ASPC..180...57R, 2017isra.book.....T}.
In this paper we describe the design, validation and commissioning of a new correlator termed ``MWAX'' for the MWA telescope.

\subsection{Legacy MWA system description}

The MWA is a low frequency (70--300\,MHz) interferometric radio telescope located at \textit{Inyarrimanha Ilgari Bundara}, the CSIRO Murchison Radio-astronomy Observatory (MRO) in Western Australia.  It was originally comprised of 128 antenna ``tiles'', where each tile is comprised of 16 dual-polarisation dipole antennas.  This original system is referred to as Phase I \citep{Tingay2013}.

Since its commissioning in early 2013, the MWA had been operating with a 128-tile signal processing system that employed a hybrid of field-programmable gate array (FPGA) and graphics processing unit (GPU) technologies.
Briefly, the original system employed a two-stage frequency channelisation scheme (a ``coarse'' polyphase filterbank (PFB) followed by a ``fine'' PFB, both implemented on FPGAs) then cross correlation and accumulation, implemented on GPUs. A full description of the original MWA correlator can be found in \cite{Ord2015}.
Here we refer to the original correlator as the MWA ``legacy correlator''.

To support high time resolution applications such as pulsar science, the legacy system also included a means to capture and store raw voltage data (after fine channelisation) from each tile, known as the Voltage Capture System (VCS) \citep{tremblay15}.

In 2017 the MWA underwent a substantial upgrade, referred to as Phase II, which included the addition of a further 128 tiles, making a total of 256 tiles \citep{Wayth2018}.  This upgrade was constrained by the available number of receivers\footnote{In the MWA context the term ``receiver'' relates specifically to a unit of electronic equipment that takes signals from multiple tiles, performs certain pre-processing functions, and outputs packetised data streams. Receiver functionality is described in more detail in \S \ref{sec:architecture}.} to capture signals, and the design of the legacy correlator to process up to 128 tiles' worth of data.  It was thus necessary to alternate the array between ``compact'' and ``extended'' configurations (to suit differing science objectives) through a sub-arraying process, where the receivers and correlator components were manually re-cabled to a different grouping of tile signals.

The signal chain for Phases I and II involved each of 128 tiles having their 16 antennas connected to an analog beamformer, the output of which was fed, along with others, to a receiver that carried out digitisation and coarse channelisation \citep{Prabu2015}.
There were a total of 16 receivers that sent voltage sample data to an FPGA-based fine channelisation stage and then into the software-based cross-correlation stage of the legacy correlator.  A good illustration of the legacy signal chain is Figure 1 of \cite{tremblay15}.

The original signal processing system of the MWA served the community well in pursuing the key science programs of the MWA \citep{2013PASA...30...31B,2019PASA...36...50B}. However the original system was inflexible, in particular the number and type of connections into and out of the fine PFB were fixed, as well as the frequency resolution of the fine PFB.
This inflexibility limited the legacy correlator to just a few observing modes, and prevented the expansion of the number of tiles that could be correlated.
The selection of tiles for the compact and extended configurations was sub-optimal for science cases requiring both sensitivity to very large scale structure whilst simultaneously resolving foreground objects (see \cite{2020PASA...37...32H} for a discussion on the impact of baseline selection for the sub-arrays).
The legacy system was also sub-optimal for astronomers wishing to conduct high time resolution science with the MWA, such that the fine PFB had to be inverted for astronomers to obtain microsecond or better time resolution required for pulsar and cosmic-ray science \citep[e.g.][]{2019ApJ...882..133K,Mcsweeney2020,2021JAI....1050003W}.

Upgrading the original MWA signal processing chain therefore required major changes, in particular removing the fine PFBs and replacing/re-purposing all systems downstream of the receivers. Using commercial-off-the-shelf (COTS) equipment where possible, and standard networking hardware and protocols, the new MWAX system is far more flexible and scalable than the legacy system.

\subsection{Key goals and design philosophy}

A primary goal in developing MWAX was to allow for future simultaneous correlation of up to and including all 256 tiles, once additional receivers were sourced.  There was also a desire to extend other aspects of functionality, to support more flexible observing modes, and to support the future expansion of the telescope.

Furthermore, the legacy correlator had been in operation since construction in 2012 up until being decommissioned in mid 2021.  By that time it was well past its design lifetime.  So another core driver was for MWAX to replace the legacy system and thus remove a serious hardware lifecycle risk.  

In developing the MWAX concept, the following considerations influenced the design philosophy:
\begin{itemize}
\item Reduce or eliminate arbitrary limits in the software and design, including removing the reliance on an inflexible hard-coded fine channelisation stage;
\item Leverage industry standard COTS hardware and software components as much as possible;
\item Reduce interdependence between components for easier support and more graceful fall-back under failure conditions;
\item Provide a flexible base for future needs and capabilities; and
\item Re-use as much as practicable existing hardware, networks and systems to reduce the up-front capital cost of the project.
\end{itemize}

Combined, these considerations result in a solution that is not only more cost effective but also easier to deploy and operate, and more straightforward to maintain and extend into the future.

\subsection{Specific requirements}

The goal with MWAX was to provide the following top-level capability enhancements over the legacy correlator:
\begin{itemize}
\item Ability to cross-correlate more tiles, up to at least 256;
\item Mitigation of operational risk through interoperation with both current and expected future receiver (digitiser) systems;
\item Ability to expand instantaneous bandwidth by means of a modular/scalable design that can support more coarse channels, should future receivers provide that;
\item Support for finer and additional frequency resolution options than just the fixed 10 kHz provided by the legacy fine PFB;
\item Support for finer and additional temporal resolution options compared to the legacy correlator;
\item Support for higher temporal resolution in voltage capture mode;
\item Support for future provision of real-time fringe stopping;
\item Improved data quality through reduced signal quantisation/ clipping within the channelisation and cross-correlation stages;
\item Increased voltage/visibility buffering space to provide gr-eater tolerance to outages of downstream data archiving facilities and networks;
\item Reduced operational cost through lower power consumption when not observing; and
\item Support for easy connectivity to other external instruments that can commensally tap into high time resolution voltage data, without impacting correlator operation.
\end{itemize}

\begin{table*}[!t]
\begin{center}
\begin{tabular}{ll}
\hline
Parameter                                 & Value \\
\hline \hline
Maximum tiles correlated                  & 256 \\
Correlation frequency resolution          & 200 Hz to 1.28 MHz \\
Correlation temporal resolution           & 250 ms to 8 seconds \\
Instantaneous bandwidth processed         & 1.28 MHz per MWAX compute node\\
Total bandwidth processed                 & 30.72 MHz for 24 MWAX compute nodes\\
Voltage capture temporal resolution       & 781.25 ns \\
Voltage capture memory buffer capacity    & 240 seconds for 128 tiles \\
\hline
\end{tabular}
\caption{Summary of Key MWAX Specifications. (Note that the design is modular and scalable:- more bandwidth can be processed by employing additional compute nodes, each processing one 1.28 MHz coarse channel.)}
\label{table:Parameters}
\end{center}
\end{table*}

The key specifications for the MWAX correlator are summarised in Table \ref{table:Parameters}.  This table lists the major parameters of the delivered system, i.e. what the implementation achieved in relation to the above design requirements.

\subsection{Architecture and technology choices}

The MWA can be categorised as a ``large-N telescope'', where N is the number of antenna elements.
The 128 dual polarisation tiles results in 8,256 distinct tile pairings (baselines), with four cross-polarisation cases per baseline.
The majority of use-cases require visibilities to be produced over a number of fine frequency channels, typically with width in the order of kHz to tens of kHz.
For low frequency, relatively compact arrays like the MWA, correlation is commutative in that it is equally possible to correlate at the full coarse channel resolution and then fine channelise the visibilities, or fine channelise the coarse channel voltage time-series and correlate each individual fine channel.  The former is referred to as the ``XF'' approach and the latter the ``FX'' approach (see \S \ref{sec:architecture}).  For large-N telescopes the FX approach is generally favoured because it results in fewer overall mathematical computations, due to there being fewer voltage time-series than baselines to channelise.

The rapid advancement in compute capability and power efficiency of general purpose GPU technology has facilitated development of software correlators that are now competitive in power consumption with hardware solutions (utilising FPGAs or application-specific integrated circuits (ASICs)).  At the same time, software correlator solutions are generally less costly in development, are of greater flexibility in operation, and provide improved scalability/extendability and maintainability.  With the GPU approach it is straightforward to perform a hardware refresh simply by upgrading the GPU platform with little or no change to software.  For example, a future addition of more tiles beyond the MWAX design requirement of 256 should not necessitate a redevelopment of the correlator: the correlation of all tiles may be accommodated by simply upgrading to newer generation GPUs, most likely accompanied by a reduction in power consumption (see \S \ref{sec:SKA}).


Other large-N telescopes have pursued the path of software correlation using GPU-accelerated COTS compute hardware. In a number of cases a hybrid approach has been adopted where the F stage of an FX correlator has been implemented with FPGAs and the X stage in GPU software.  Examples include the correlators for the Large Aperture Experiment to Detect the Dark Ages \citep[LEDA;][]{2015JAI.....450003K}, the Canadian Hydrogen Intensity Mapping Experiment \citep[CHIME;][]{Denman2015AGC}, the upgraded Molonglo Observatory Synthesis Telescope \citep[UTMOST;][]{Bailes2017a},
the upgraded LOFAR correlator COBALT-2 \citep{2018A&C....23..180B},
and the legacy MWA correlator that MWAX is replacing \citep{Ord2015}. 
In the case of \cite{Bailes2017a} the correlator system employs commmodity ethernet networking with a multicast architecture to facilitate commensal processing beyond correlation, such as real-time beamforming, calibration, radio frequency interference (RFI) mitigation, transient detection, etc.

These trends were a major influence on the philosophy taken with MWAX.
Note that with MWAX we have taken the use of GPUs a step further and implemented both the second F and X stages in GPU software, for the reasons explained in \S \ref{sec:f_engine}.

Following the design description we present a summary of the MWAX validation and commissioning activities (\S \ref{sec:validation}), followed by a road-map for its potential future extensions and enhancements (\S \ref{sec:future}), and a summary of the overall outcomes and conclusions of the work (\S \ref{sec:summary}).

\section{MWAX DESIGN}
\label{sec:design}

\subsection{Top-level architecture} \label{sec:architecture}

\begin{figure*}[t]
\begin{center}
\includegraphics[width=450pt]{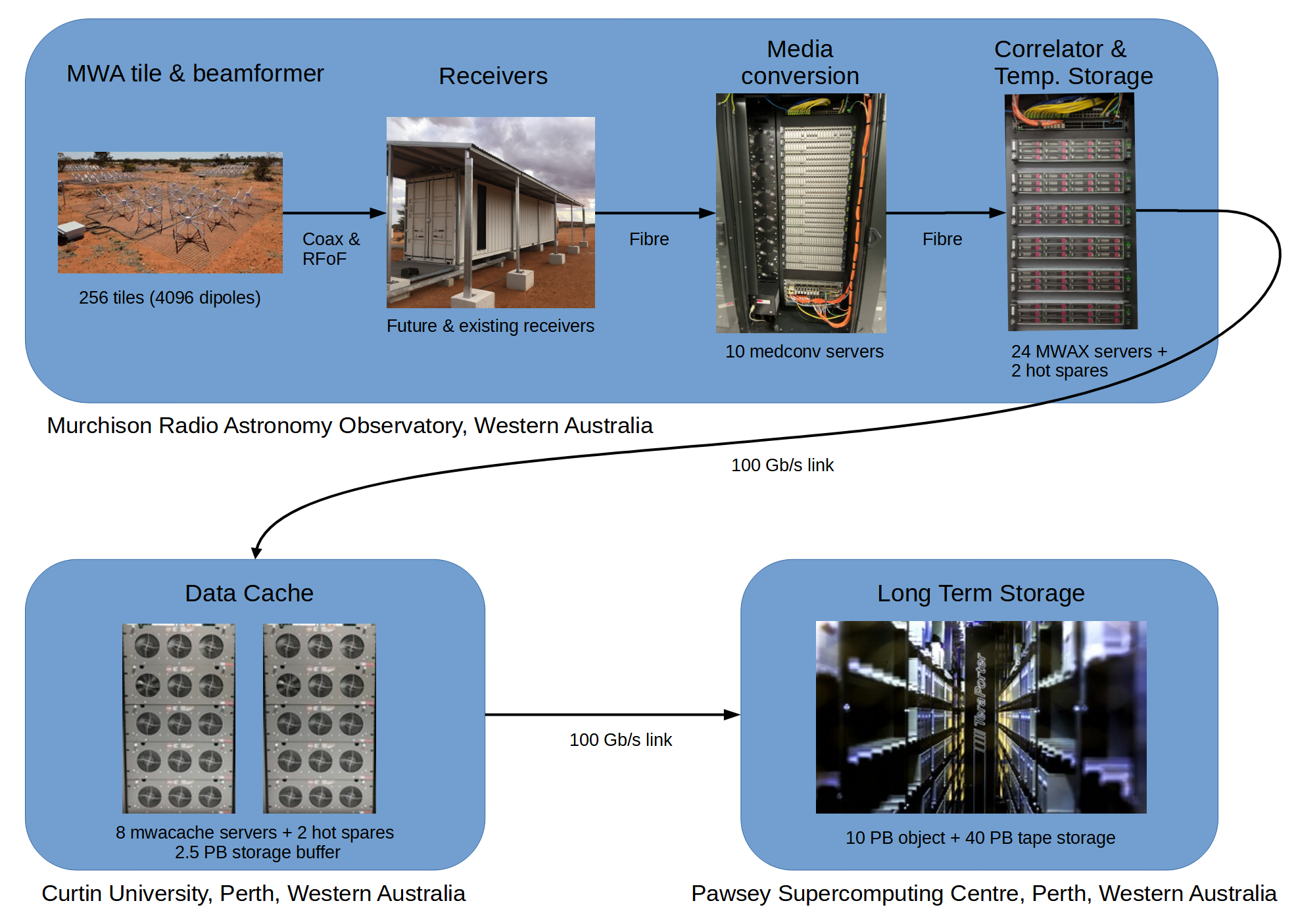}
\caption{MWA components and data flow.}
\label{fig:MWA_architecture}
\end{center}
\end{figure*}

\begin{figure*}[t]
\begin{center}
\includegraphics[width=440pt]{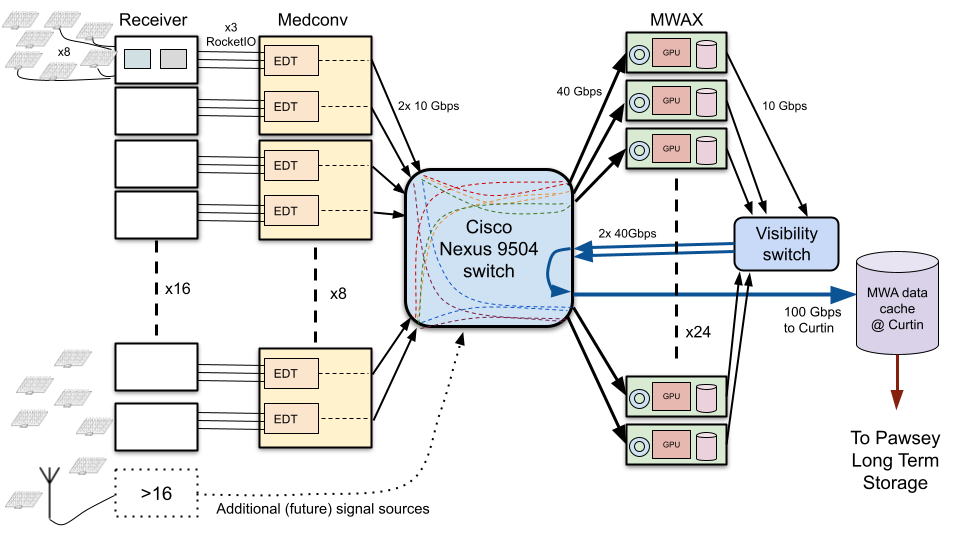}
\caption{MWAX components and signal paths. The coloured dashed lines within the Nexus 9504 switch represent the routing of data performed by the switch, such that all signals for a specific frequency channel are routed to a single MWAX Server.}
\label{fig:MWAX_signal_path}
\end{center}
\end{figure*}

The major components and end-to-end data flow of the MWA are illustrated in Figure \ref{fig:MWA_architecture}.
At the heart of the MWAX correlator is a set of GPU-enabled compute nodes, each of which processes a fraction of the total signal bandwidth.
These nodes are preceded by digital signal acquisition and conversion systems 
and followed by temporary storage and transmission of visibility data to the long term storage.

The MWAX components and signal flow between them are further decomposed in Figure \ref{fig:MWAX_signal_path}.
There are 16 existing receivers in the field, with a 17\textsuperscript{th} and 18\textsuperscript{th} currently being commissioned.  Each receiver accepts dual-polarisation analog-beamformed signals from 8 tiles.  As described by \cite{Prabu2015}, the receivers digitise 327.68\,MHz of input bandwidth to a bit-width of 8 bits, and follow this by coarse channelisation into 256 channels of width 1.28\,MHz by means of an FPGA-based critically-sampled PFB \citep{crochiere}.
A subset of 24 coarse channels (total bandwidth 30.72 MHz) can be selected to send downstream. Each coarse channel is quantised to 5-bit complex samples and output on 3 fibre optic cables per receiver using the proprietary Xilinx RocketIO protocol\footnote{RocketIO is a data transport protocol used by many Xilinx FPGA products. See: https://www.xilinx.com/}.  The fibre optic cables terminate in the control building at the media conversion servers. 
These capture the RocketIO data and convert it into Ethernet User Datagram Protocol (UDP) packets, which are sent out to a Cisco Nexus 9504 switch as multicast data.
Future generation receivers, including the 17\textsuperscript{th} and 18\textsuperscript{th} now being commissioned,
will packetise the output data directly to MWAX-compatible multicast UDP format and connect directly to the Nexus switch.

The ``MWAX Servers'' are 24 GPU-accelerated rack server computer nodes.  They act as multicast clients that accept those UDP packets addressed to them and assemble data for a given coarse channel into segments of 8 seconds, known as sub-observations (see \S \ref{sec:udp_capture}).
By design, each MWAX Server processes only a single coarse channel, hence the major corner-turning operation is provided automatically by the switch.
The raw data rate of UDP packets ingested by each MWAX Server is a function of the number of tiles in the current array configuration.  For N tiles, the average ingest rate is N*5.14 MB/s, which for 128 tiles equates to 660 MB/s and for 256 tiles equates to 1.32 GB/s.  These rates are comfortably within the capacity of the 40 GbE input link to the server, but there are other constraints around memory size, hard disk write speed and internal bus traffic that reinforce the decision to limit each MWAX Server to a single coarse channel.

Within an MWAX Server each sub-observation data block for a coarse channel is fine channelised, cross-correlated, and frequency/time averaged (see \S \ref{sec:fx_engine}).  The cluster of 24 MWAX Servers is capable of simultaneous correlation and buffer storage for up to 256 tiles and up to 24 coarse channels.
%

\begin{figure*}[t]
\begin{center}
\includegraphics[width=\textwidth]{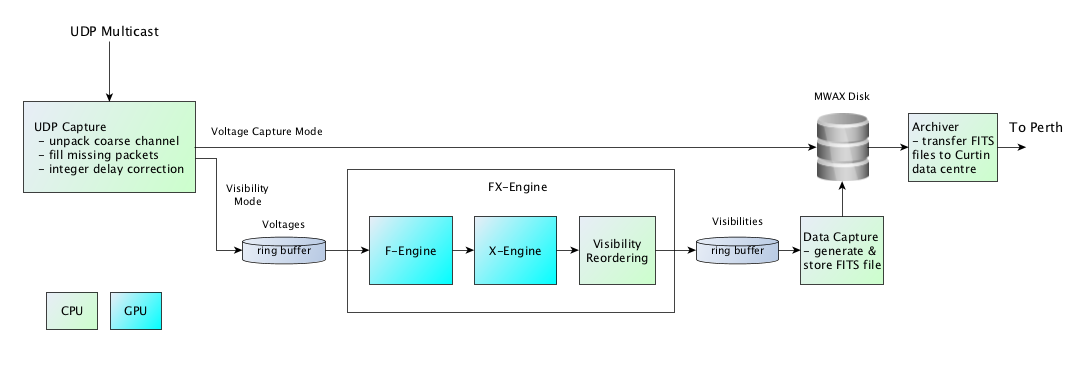}
\caption{The functional components of an MWAX Server, showing the partitioning between CPU and GPU implementation.}
\label{fig:MWAX_server}
\end{center}
\end{figure*}

The key design parameters of MWAX are summarised in Table \ref{table:Parameters}.
Note that not all combinations of frequency and temporal resolution are simultaneously available due to real-time processing constraints and/or the data volume of visibilities generated.  Appendix 1 provides a table of the available modes when correlating 128 tiles.  For 256 tiles the modes available will be further constrained.

Table \ref{table:Parameters} also provides specifications for the temporal resolution and buffer capacity for voltage capture mode with 128 tiles. 

MWAX's multicast architecture allows all receiver units\footnote{The existing receiver units require media conversion to create multicast UDP packets.} to each send their streams of high time resolution data to any number of multicast consumers with no additional load on the sender.  At the time of writing there are 24 multicast consumers (one for each coarse channel for correlation and voltage capture), however, the multicast architecture allows for additional consumers to utilise the same high time resolution data for other purposes, e.g. RFI monitoring equipment or transient detection pipelines.
Such applications can commensally consume some or all of the captured high time resolution data without impacting the operation of the correlator.


Each MWAX Server implements the functions shown in Figure \ref{fig:MWAX_server}.
The real-time data flows on the MWAX Servers are managed through the use of input and output ring buffers that decouple its computational workflow from the input source and output destinations.  These ring buffers are established and accessed using the open-source ring buffer library PSRDADA \citep{psrdada2021}.

MWAX employs the “FX” correlation architecture where the input time samples for each signal path (antenna and polarisation) are fine-channelised prior to cross-correlation, reducing the correlation process to complex multiplications in the frequency domain  \citep[][Ch 8]{2017isra.book.....T}.
In this architecture the first stage of processing (channelisation) is performed by the ``F-engine''.  The MWAX F-engine is described in more detail in \S \ref{sec:f_engine}.  The second stage of processing (cross-correlation) is performed by the ``X-engine''.  For the MWAX X-engine, development time/cost was minimised by using the existing open-source GPU correlator library ``xGPU'' \citep{Clark2013} (the same library that was at the heart of the legacy correlator). 

Time and frequency averaged visibility data are transferred from temporary storage on site, through the MWA data cache servers at Curtin University, and on to the MWA's long term data archive hosted by the Pawsey Supercomputing Centre.

\subsection{Server hardware}  \label{sec:hardware_networking}

As defined previously, the core hardware component is a cluster of GPU-accelerated rack server compute nodes -- the ``MWAX Servers''.  There are currently 24 active nodes plus two hot spares.  There are ten additional rack servers performing media conversion (that were re-purposed from the legacy correlator) and an assortment of switches, optics, and other equipment.  All together the MWAX system occupies three of the MWA's 19" racks on site.

The MWAX Servers are split across two 42 RU racks in the Control Building at the MRO. Each rack is fitted with a managed power distribution unit and a Dell N1548 managed switch allowing remote management, monitoring, and power cycling of all servers. 

Each MWAX Server node has the following specification:
\begin{itemize}
\item Model: HPe Proliant DL385 Gen10+ 2RU rack-mounted server;
\item CPU: 2 x AMD Epyc 7F72 CPUs (24 cores per socket);
\item Memory: 512 GB (16 x 32 GB Dual Rank PC4-3200 MHz DRAM);
\item Boot disks: 2 x 480 GB SATA SSD disks in RAID1;
\item Data disks: 12 x 6 TB SAS 7.2K rpm disks in RAID5;
\item RAID Controller: HP Smart Array P816i-a SR with 4 GB battery-backed cache;
\item GPU: NVIDIA A40 GPU (48 GB RAM on board);
\item Network: Intel i350 1GbE network adpater OCP3 adapter;
\item Network: Mellanox MCX516A-BDAT 40GbE PCI adapter;
\item Network: HP/Mellanox MCX512F 10/25GbE PCI adapter; and
\item Power Supply: dual-redundant 1600 W power supply units (PSU).
\end{itemize}

The MWAX Servers boot from the 2 x 480 GB SSD disks which are configured in RAID 1 (mirrored disks). Ubuntu Server 20.04 LTS is installed on each server, and then Ansible\footnote{Ansible is an open source IT automation tool which allows scripting of server installations and configurations. See: https://www.ansible.com/} automated deployment scripts are used to deploy packages, drivers, and MWAX software.
The 12 x 6 TB SATA disks are configured in a RAID5 volume, split into a ``voltdata'' partition of 30 TB and a ``visdata'' partition also of 30 TB. Voltage capture data is written to the voltdata partition at up to 2 GB per second.
This write speed is possible due to having 12 disks in a parallel XFS filesystem, and that the voltdata partition is the first partition on the disk. The second partition, visdata, occupies the slower half of the RAID volume and is used for writing visibility data, which has much more modest disk write speed requirements of less than 1 GB per second at the highest resolution correlator modes.

The servers allow interactive login and communication with the Monitoring and Control (M\&C) system and Ansible deployments via the 1GbE interface connected via unshielded twisted pair (UTP) ethernet cable. MWAX uses one port of the 40GbE Mellanox network adapter to ingest the MWAX multicast UDP data from the Media Converter servers (or future new receivers). One of the Mellanox 10GbE ports is for archiving visibility and voltage data to the mwacache servers in Perth. The 40GbE network connection terminates at the Cisco Nexus 9504 switch via an active optic cable (AOC).
The 10GbE network connection is terminated at a Fibrestore N5860-48SC aggregation switch which has a 40GbE backhaul to the Cisco Nexus 9504 switch. An aggregation switch was implemented to ensure there are ample free 10GbE ports on the Cisco Nexus 9504 switch for future receiver requirements.

The traffic to Perth for visibilities and voltages uses the repeat-request Transmission Control Protocol (TCP), so there is no risk of introducing packet loss that might occur if using UDP.  In a near future upgrade, a second 40GbE backhaul link will be established to provide 80 Gbps bandwidth between the aggregation switch and the Cisco Nexus 9504 switch. The Cisco Nexus 9504 switch also hosts the 100GbE link to Perth.

\subsection{MWAX media conversion} \label{sec:medconv}

The native MWAX ingest multicast UDP data format is suitable for direct generation by modern receivers fitted with Ethernet compatible ports.  This data format can be transported and routed over COTS data links/switches and routers.  The existing 16 legacy MWA receivers, however, cannot support this format directly.
For this reason, while MWA continues to use these receivers for digitisation and coarse channelisation, it is necessary to provide a conversion stage for compatibility between the legacy output data format and the Ethernet MWAX format.

This media conversion (``medconv'') stage consists of ten x86 servers selected and re-tasked from the pool of 16 ``VCS Servers'' freed during the legacy correlator decommissioning.  Each server is fitted with two Engineering Design Team Incorporated (EDT)\footnote{See: https://edt.com/} Xilinx FPGA PCIe cards capable of ingesting the RocketIO low layer protocol of the older format.  Custom software on the servers takes these RocketIO packets from the receivers in groups of 2048 and outputs 128 Ethernet multicast UDP packets in MWAX format.
The RocketIO packet format contains 1 time sample, for 8 coarse channels, for 16 radio frequency (RF) signals (8 dual polarisation tiles). The output packets contain 2048 time samples, for 1 coarse channel, for 1 RF signal.

The MWAX format data are then addressed via multicast to a coarse channel multicast group and sent to the main Cisco Nexus 9504 switch via 10GbE Ethernet, where they join compatible MWAX voltage data packets coming directly from new receivers.  This allows cross correlation between heterogeneous receiver hardware.

\begin{figure}[b]
\begin{center}
\includegraphics[width=223pt]{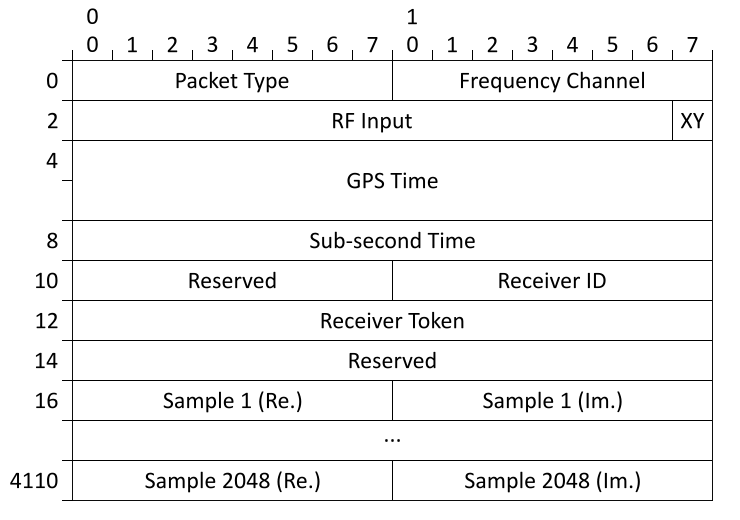}
\caption{UDP packet structure for coarse channel data.}
\label{fig:MWAX_udp_packet_structure}
\end{center}
\end{figure}

MWAX data packets carry several information fields in addition to raw voltage data, including identifiers for the packet type, RF source and polarisation, and the receiver from which the packet was generated. Figure \ref{fig:MWAX_udp_packet_structure} shows the binary format of these packets. Each packet is also marked with a timestamp comprising a 32-bit GPS time (in seconds) and a 16-bit subsecond time, measured in number of packets since the start of the 1 second GPS time window.

\subsection{MWAX UDP and voltage capture} \label{sec:udp_capture}

As per standard IPv4 multicast, any device on the voltage network can “join” the multicast group for one or more coarse channels and a copy of the relevant stream will be passed to them.  Each MWAX Server executes a process to perform UDP capture for a single coarse channel.  This process assembles packets from the multicast stream in shared memory (RAM) into 8 second blocks of high time resolution voltage data based on their time and source.  Each 8 second block is known as a ``sub-observation''. At the completion of each sub-observation, the RAM file is closed and made available to another process.  The file is referred to as a ``subfile''.  Depending on the current observing mode, the subfile may be:
\begin{itemize}
\item Retained in RAM for a period to satisfy triggered ``buffer dump'' commands;
\item Written immediately to disk for voltage capture mode; or
\item Passed to the FX-engine for cross-correlation via a PSRDADA ring buffer.
\end{itemize}
A 256 tile sub-observation buffer for one coarse channel, for the 8 seconds, is approximately 10 GB in size. An MWAX Server can buffer approximately 4 minutes of data in its available memory for 128 tiles.

As each 8 second subfile is created in the /dev/shm directory (shared memory filesystem), another process reads the header. If the MODE key/value is \texttt{MWAX\_VCS} a sub-process copies the subfile, using the standard Linux cp command, to the RAID 5 /voltdata partition. Alternatively, if the MODE is \texttt{MWAX\_CORRELATOR} then the subfile is transferred into the input PSRDADA ring buffer for correlator processing. If any other mode value is present (for example \texttt{NO\_CAPTURE}, which signifies no current observation in the schedule) then no action is taken. In any case, once the MODE has been read and acted upon, the subfile is renamed to become a``.free'' file and is then able to be re-used by the UDP Capture process.


\subsubsection{Integer delay correction} \label{sec:int_delay}

The UDP data from the media converters are timestamped based on the timestamps from the receivers. The MWA system contains several sources of static time delays, including non-equal optical fibre lengths carrying clock signals, non-equal coaxial cable lengths carrying RF signals, and for the newer long baseline tiles, long optical fibre runs (of order kilometres) also carrying RF signals. The legacy system partially corrected for some static delays pre-correlation, but by design tolerated these delays (as well as geometric delays associated with tracking an astronomical source) with 10\,kHz frequency resolution and corrected for them post correlation \citep{2015PASA...32....8O}.
This system worked, but placed significant constraints on observing modes to avoid bandwidth and temporal decorrelation \citep{Wayth2018}, resulting in very large data rates for MWA Phase II extended array observations.

MWAX UDP Capture can optionally apply the composite delay required for each RF signal to compensate for fixed and astronomical geometric delays.  It then selects an integer number of samples to advance or delay the data stream before  writing to the subfile.  The residual fractional sample components of the required delays are recorded in the metadata block (block 0) of the subfile where they can be accessed by the F-stage of the correlator for fractional delay correction through phase rotations (see \S \ref{sec:f_engine}). The whole sample signal shifts applied to each signal path remain static across the entire subfile. The fractional components for each signal path are updated with a 5 ms cadence, hence there are 1,600 unique values per signal path placed in the metadata block to cover the entire 8 seconds of the sub-observation.

\subsubsection{Minimising multicast UDP packet loss}

The UDP Capture process also deals with any lost or excised packets by setting the missing data to zero and setting ``zeroing flags'' in the metadata block of the subfile to indicate, with 50 ms resolution, which signal paths of which sub-blocks were zeroed out.  This information can be used by the correlator to calculate what fraction of each integration time, for each baseline, contained valid data, and to normalise the output visibilities accordingly (see \S \ref{sec:vis-weights}).

Maintaining zero or very low packet loss when receiving high data rate multicast UDP requires careful tuning of the Linux kernel, network, and software. The process, techniques, and rationale for optimising and testing network throughput and lowering packet loss is an extensive topic in its own right and highly system/situation dependent and is out of the scope of this paper.  However, the MWAX development team did identify some elements that were essential to the process, which are described below.

The Linux kernel has many parameters associated with networking configuration that can be modified via the ``sysctl'' tool. We spent considerable time optimising sysctl values and found that in most cases the defaults needed to be increased in order to attain higher packet throughput with minimal loss. The \texttt{rp\_filter} settings refer to reverse path filtering, and this had to be disabled (set to zero) for the 40 GbE network interface to receive any multicast UDP packets.  Specifically, we optimised the following parameters:

\begin{itemize}
\item \texttt{net.core.rmem\_max};
\item \texttt{net.core.wmem\_max};
\item \texttt{net.core.rmem\_default};
\item \texttt{net.core.wmem\_default};
\item \texttt{net.core.netdev\_max\_backlog};
\item \texttt{net.core.netdev\_budget};
\item \texttt{net.core.optmem\_max};
\item \texttt{net.ipv4.conf.default.rp\_filter}; and
\item \texttt{net.ipv4.conf.all.rp\_filter}.
\end{itemize}

The MWAX UDP capture process is intentionally run on the same NUMA node
that the Mellanox 40 GbE network card is allocated. This prevents transfer of data across the inter-node bus (AMD Infinity Fabric), which would otherwise result in increased latency and lead to lower throughput and dropped packets. Similarly, the mapping of interrupts to cores for the Mellanox 40 GbE network card is locked to specific cores on the same NUMA node. An additional optimisation was to use the Linux ``ethtool'' to increase the receive ring buffer size to the maximum allowed by the Mellanox driver.  

\subsection{FX-engine} \label{sec:fx_engine}

The MWAX and legacy correlators both employ the “FX” architecture.  However, whereas the legacy correlator utilised a PFB \citep{crochiere} for its fine channel F-engine, MWAX employs the Fast Fourier Transform (FFT) \citep{Bracewell1999}.  The approach involves over-resolving the input signals in the frequency domain to create a large number of “ultra-fine” channels.  These channels are at a resolution that is finer than needed to accomplish cross-correlation in the frequency domain, but the data volume into the X-engine is the same in terms of total time-frequency samples, which is what drives the X-engine compute complexity.  At the output of the X-engine are ultra-finely-channelised visibilities, which are subsequently grouped (averaged) to provide the final desired output visibility channel width\footnote{Channel averaging following correlation to obtain a specific fine channel width is equivalent to grouping the ultra-fine channels into that same width prior to cross-correlation by means of multiple inverse-FFT (IFFT) processes (one per fine channel).  That method is sometimes referred to within the astronomy domain as a ``convolving filterbank'' \citep{vb11}.  The MWAX approach of combining channels post-correlation means that computationally expensive IFFT operations are replaced with simple summations.}.
Averaging over a large number of ultra-fine channels creates final output channels with a relatively flat response and small spectral leakage between channels – effectively emulating what is accomplished had one used a PFB channeliser (see \S \ref{sec:averaging}).  However, there are two distinct advantages in using the FFT/averaging approach over a PFB for the F-engine:
\begin{enumerate}
\item Simple and efficient GPU implementation using highly-optimised standard NVIDIA FFT library code ``cuFFT''\footnote{https://developer.nvidia.com/cufft}, which is preferable to a costly and inflexible (typically FPGA-based) PFB solution.  With a GPU implementation (using floating point arithmetic) there are not the same scaling/overflow issues as when implementing FFTs with FPGAs (which use integer arithmetic), which means large FFTs can be employed, thus facilitating the ultra-fine channel approach; and
\item Increased flexibility in both the input ultra-fine channel resolution and the output visibility channel resolution. If desired, much finer output visibility channelisation is available than the PFB approach (at the expense of increased spectral leakage), or very low spectral leakage is available by choosing wider output channels (or any trade-off in between).
\end{enumerate}

\begin{figure*}[ht]
\begin{center}
\includegraphics[width=498pt]{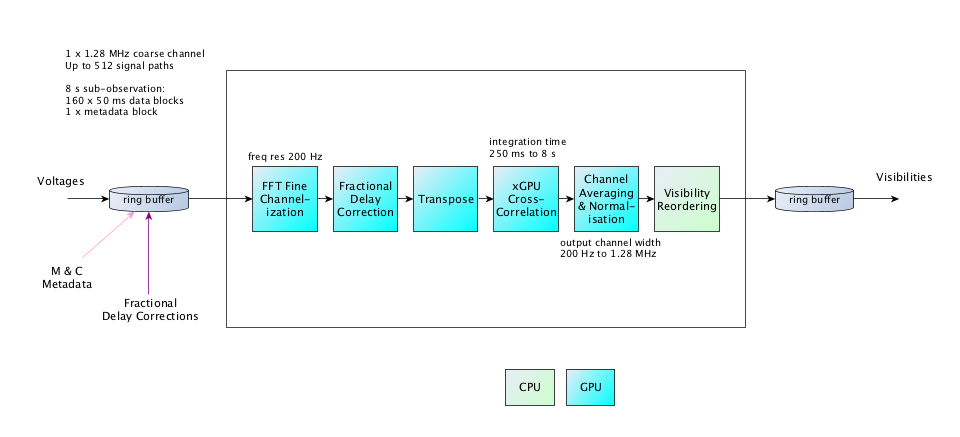}
\caption{The functions of the MWAX FX-engine.  Voltage data in the form of 8 second subfiles are loaded into a ring buffer, along with M\&C and delay correction metadata.  They are processed by a pipeline of functions on the GPU before the CPU places the output visibilites into the desired order and writes them into an output ring buffer.}
\label{fig:MWAX_FX_engine}
\end{center}
\end{figure*}

The MWAX correlator FX-engine is implemented as a PSRDADA client; a single process that reads/writes from/to the input/output ring buffers, while working in a closely coupled manner with a single GPU device - which can be any standard NVIDIA/CUDA-based GPU card.
Figure \ref{fig:MWAX_FX_engine} shows the processing stages and data flows within the MWAX correlator FX-engine process.

The FX-engine treats individual 8 second subfiles as independent work units. Most of its mode settings are able to change on-the-fly from one subfile to the next.  It nominally operates on 50 ms units of input voltage data, i.e. a total of 160 blocks over each 8 second sub-observation.
An additional block of metadata (of the same size as a 50\,ms data block) is prepended to the data blocks, making a total of 161 blocks per sub-observation.  At the start of each new sub-observation, the metadata block is parsed to configure the operating parameters for the following 160 data blocks.

The coarse channel sample rate is 1,280,000 samples per second.  Each 50 ms data block consists of 64,000 time samples for each signal path presented at the input (number of tiles x 2 polarisations).  The 256 tile correlator configuration supports up to 512 signal paths.

\subsubsection{F-engine} \label{sec:f_engine}

The conversion of coarsely channelised time series voltage data to the spectral domain takes place in the F-engine.  Data are processed one 50 ms block at a time, repeated 160 times.  The data for a block is first transferred to GPU memory where they are promoted from 8-bit integers to 32-bit floats.  The 64,000 complex time samples of each path are partitioned into 10 sub-blocks of 6,400 complex samples (representing 5 ms of time), each of which is transformed with a 6,400-point FFT on the GPU using the cuFFT library function.  This results in 10 time samples on each of 6,400 ultra-fine channels of resolution 200 Hz.

MWAX can optionally apply delay corrections to each signal path prior to cross-correlation. This will allow the future roll-out of the capability to establish a specific correlation pointing centre and optionally track it as the Earth rotates, known as ``fringe stopping'' (see \S \ref{sec:delay-correction}). These corrections are accomplished in two parts: whole-sample delays can be applied as subfiles are being assembled (see \S \ref{sec:int_delay}), and any remaining fractional sample delay components can be applied by multiplying the frequency-domain samples of each signal path by a phase gradient. The desired fractional delay values, with a 5 ms update rate, are generated externally to the FX-engine and passed via the prepended metadata block of the input subfile. A different phase gradient is typically applied to each signal path, where the gradients can differ in both their slope and phase offset.  The slopes depend only on the delays to be applied, while the phase offsets must take account of the sky frequency of the current coarse channel being processed. For computational efficiency the complex gains of the required phase gradient corresponding to the desired delay are taken from a pre-computed look-up table. The complex frequency-domain data samples are multiplied by the complex gains of the look-up table gradient and then again by the complex gains corresponding to the required phase offset.

The delay/phase-corrected frequency-domain data are then transposed to place them in the order that the X-engine requires (slowest-to-fastest changing): [time][channel][tile] [polarisation].

\subsubsection{X-engine} \label{sec:x_engine}

The MWAX X-engine uses the existing open-source GPU correlator library xGPU \citep{Clark2013}, the same library as used in the legacy correlator.  A minor but crucial modification was made to the standard xGPU library code, altering the way in which input data are ingested\footnote{https://github.com/MWATelescope/mwax-xGPU.git}.
By default, xGPU fetches its input data from host computer memory, which involves transfers from host memory to GPU memory via the host’s PCI bus.  Since MWAX’s F-engine is also implemented on the GPU, the channelised data would first need be written back to host memory before xGPU fetches them, adding further to the PCI bus traffic.  For a 256 tile correlator this two-way bus traffic seriously impacts the overall system speed and prevents real time operation.  For MWAX this traffic is completely eliminated by having the F-engine write its results directly into xGPU’s input holding buffer in GPU memory and disabling the normal fetches from host memory.  In addition, xGPU’s method of returning visibility data to the host is also bypassed.  Instead, the channel averaging is performed on xGPU’s raw output while it is still in GPU memory.  Only after averaging has been performed is the much-reduced data volume transferred back to the host.

The F-engine writes its transposed output data directly into xGPU’s input holding buffer as the data are being re-ordered.  The data from five 50 ms blocks are aggregated in this buffer, corresponding to an xGPU ``gulp size'' of 250 ms.  The minimum integration time is one gulp, i.e. 250 ms.  The integration time can be any multiple of 250 ms for which there are an integer number of such integration times over the full 8 second sub-observation (see Appendix 1).

The cross-correlation operations performed by xGPU are documented in detail in \cite{Clark2013} and \cite{Ord2015}.

\subsection{Channel averaging}
\label{sec:averaging}

xGPU places the computed visibilities for each baseline, with 200 Hz resolution (6,400 channels), in GPU memory.  A GPU function then performs channel averaging according to the \texttt{fscrunch} factor specified in the PSRDADA header of the input data file.  This reduces the number of output channels to (6400/\texttt{fscrunch}), each of width (200*\texttt{fscrunch}) Hz. The averaged output channel data are then transferred back to host memory.

During this averaging process, each visibility can have a multiplicative weight applied, based on a data occupancy metric that takes account of any input data blocks that were missing due to lost UDP packets or RFI excision (a potential future enhancement).

Ultra-fine channels are ``centre symmetric'' in relation to the coarse channel bandwidth, i.e. there is a centre ultra-fine channel whose centre frequency coincides with the centre (DC) frequency of the coarse channel.
That centre ultra-fine channel is excluded when forming the centre output visibility fine channel to eliminate any residual DC offset present in the coarse channel signal. The output values for the centre fine channel are re-scaled accordingly.
Note that only 200 Hz of bandwidth is lost in this process, rather than a complete output channel as was the case with the legacy correlator.

The output fine visibility channels are also arranged centre symmetrically across the coarse channel, i.e. there is a centre fine channel whose centre frequency coincides with the centre (DC) frequency of the coarse channel.  Adjacent channels extend symmetrically above and below that centre fine channel.  Where there is an odd number of fine channels across the coarse channel, the outermost edges of the outermost fine channels will coincide with the edges of the coarse channel.  Where there is an even number of fine channels across the coarse channel, there will be half-channels at the edges, which should be ignored.

When averaging ultra-fine channels, it matters whether the \texttt{fscrunch} factor is odd or even.  When \texttt{fscrunch} is odd it is straighforward to maintain symmetry because, over the bandwidth making up each output fine channel, there will be an equal number of ultra-fine channels above and below the centre ultra-fine channel.  When \texttt{fscrunch} is even, symmetry is maintained by averaging over (\texttt{fscrunch} + 1) ultra-fine channels, where the outermost ultra-fine channels are each weighted by 0.5.

\begin{figure}[t]
\begin{center}
\includegraphics[width=220pt]{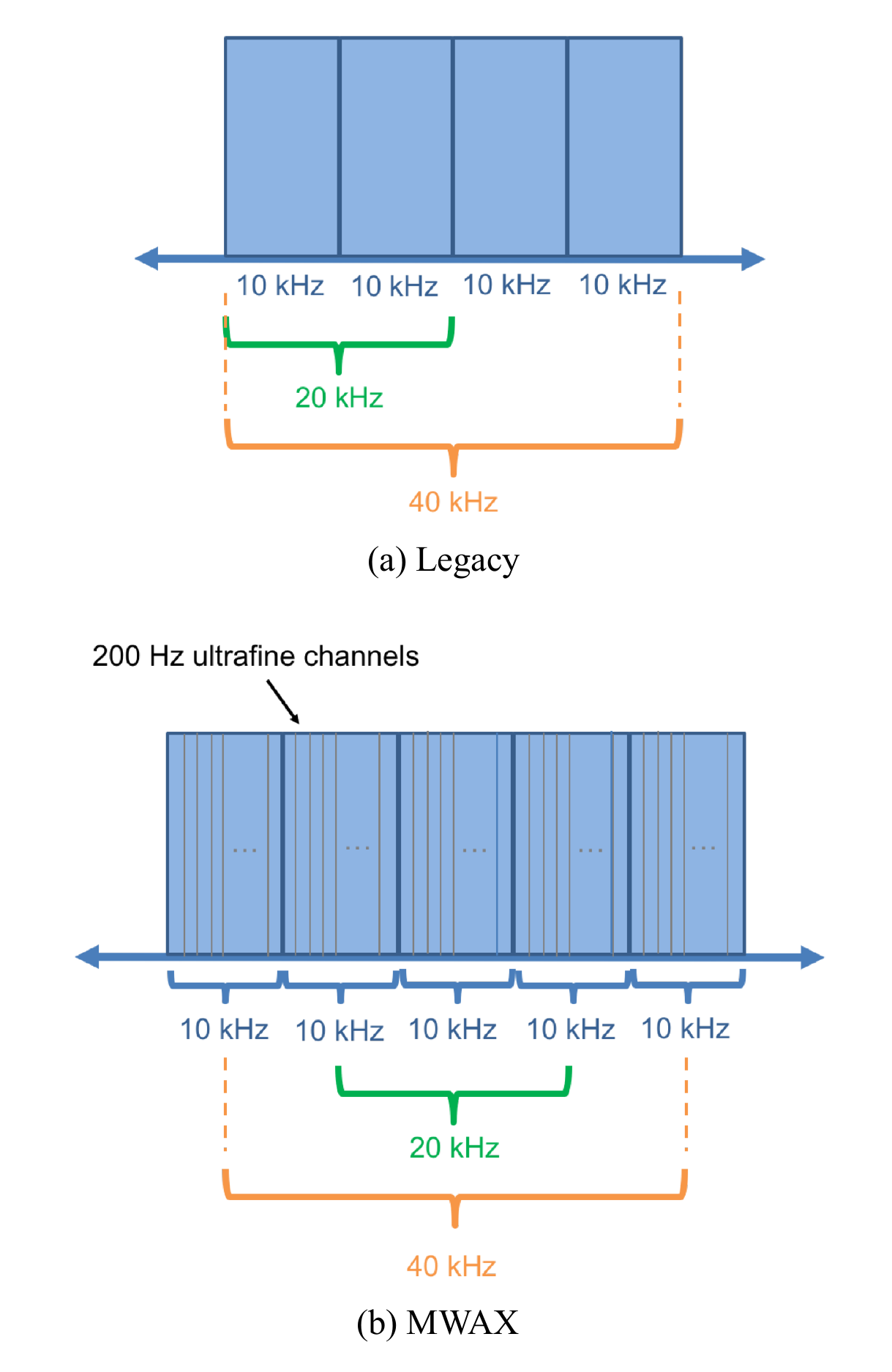}
\caption{Fine channelisation and channel averaging. (a) legacy correlator. (b) MWAX correlator.}
\label{fig:MWAX_channel_averaging_vertical}
\end{center}
\end{figure}

Since the legacy fine PFB produced only 10\,kHz output fine channels, when downstream tools performed channel averaging to obtain 20\,kHz or 40\,kHz channel widths, this was accomplished by grouping whole sets of 10\,kHz channels, as illustrated in Figure \ref{fig:MWAX_channel_averaging_vertical}(a).
This resulted in an offset between the centres of 20\,or 40\,kHz channels and the centre of the 10\,kHz channels (and the centre of the coarse channel).

MWAX always assembles its fine channels from ultra-fine channels in a centre symmetric manner, as illustrated in Figure \ref{fig:MWAX_channel_averaging_vertical}(b) for the case of 10\,kHz output channels (\texttt{fscrunch} = 50).  This means that the centres of 20\,and 40\,kHz (and all other width) channels remain aligned with the centre of the coarse channel.

The ultra-fine channels that are averaged to produce output fine channels are themselves produced by an FFT and so, prior to any averaging, the frequency response displays the classic ``scalloping'' amplitude ripple of an FFT \citep{Bracewell1999}.
The averaging process acts to overcome the scalloping and flatten the frequency response within an output fine channel.  The larger the \texttt{fscrunch} factor, the greater is the flattening effect.  The spectral leakage from adjacent channels is also proportionately reduced with higher \texttt{fscrunch} values.

\begin{figure*}[ht]
\begin{center}
\includegraphics[width=450pt]{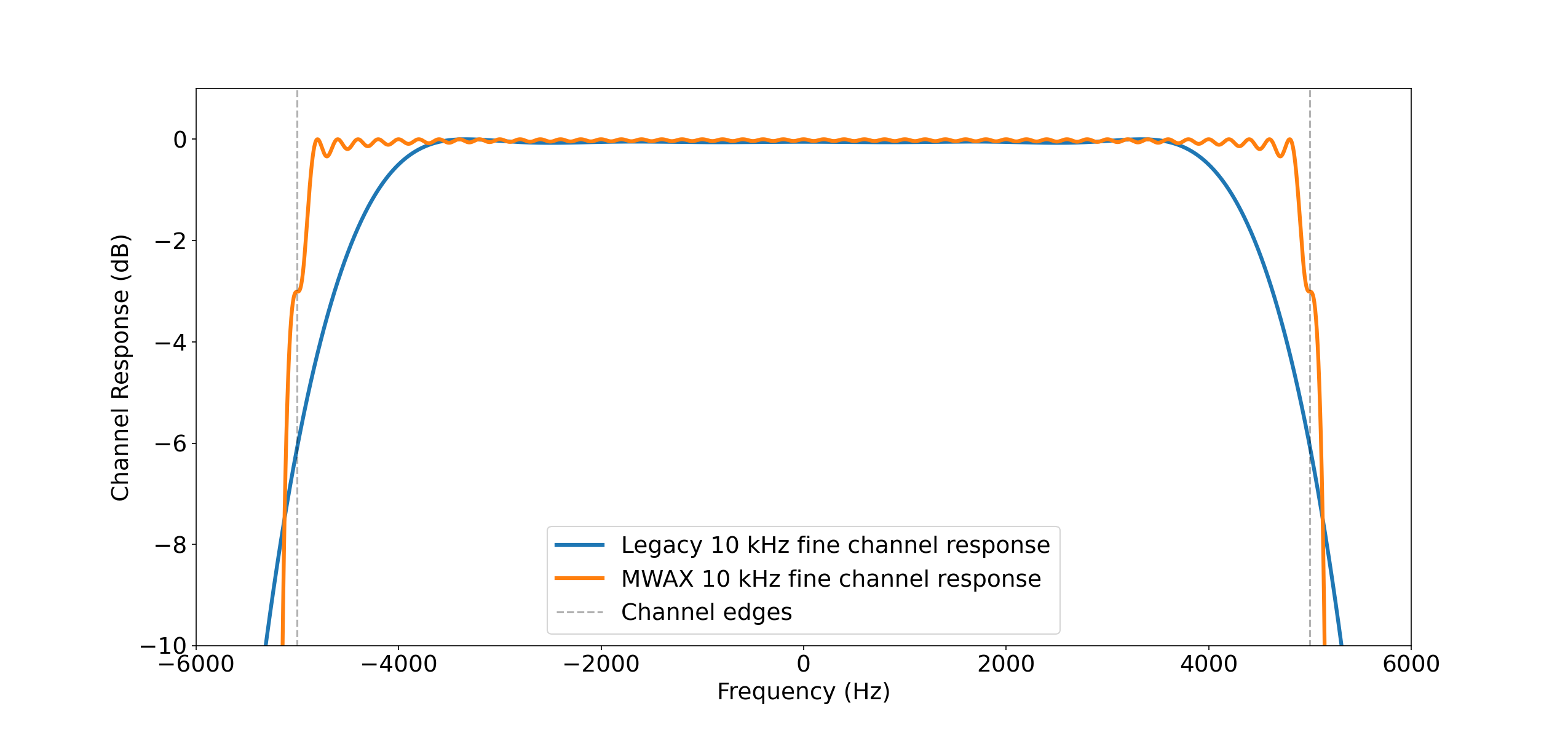}
\caption{Fine channel passband responses for the legacy and MWAX correlators, at 10 kHz fine channel bandwidth.}
\label{fig:passband_responses}
\end{center}
\end{figure*}

As an example, the resultant frequency response when \texttt{fscrunch} = 50 (10\,kHz output channels) is shown in Figure \ref{fig:passband_responses}, with the legacy fine channel response also shown for comparison.  The MWAX response extends much closer to the channel edges and exhibits steeper transitions from passband to stopband.
Furthermore, the response is precisely 3\,dB down at the channel boundaries.  Energy at the frequency of a channel boundary will appear at 50\% level in the two neighbouring channels.  Over the entire coarse channel, energy at every frequency contributes equally to the total power of visibilities. That was not the case with the legacy correlator, where deep dips exist around channel boundaries, meaning that certain sections of the coarse channel passband did not contribute equally (specifically at the edges of fine channels) to output visibilities.

Whilst forming fine channels by means of an FFT and spectral averaging provides a desirable square passband shape, stop-band rejection is inferior to a PFB solution.  This was not considered a concern for the MWA's primary use cases and was outweighed by the benefits of the chosen approach, in particular the flexibility it offers in the output fine channel width.

\begin{figure*}[t]
\begin{center}
\includegraphics[width=400pt]{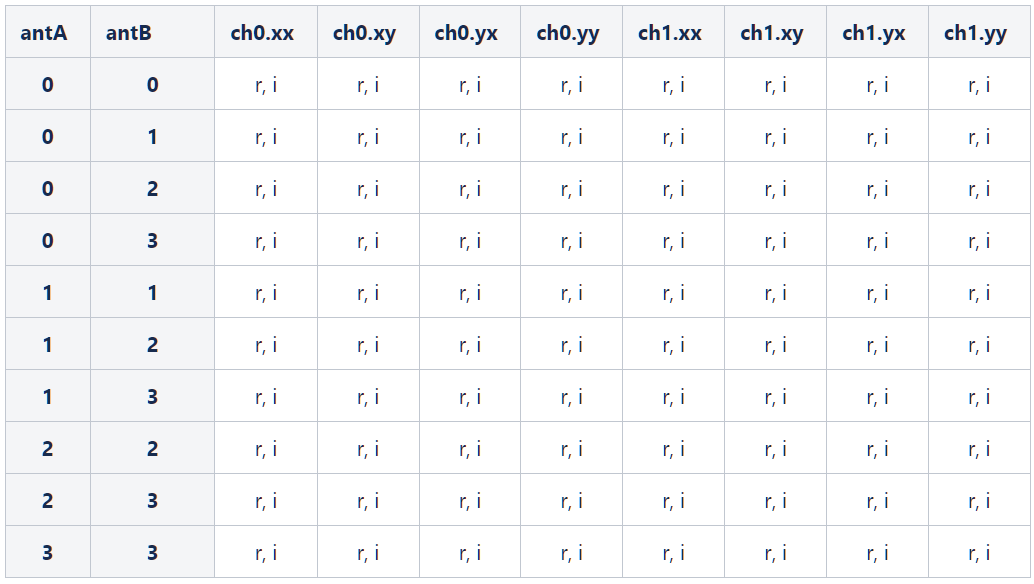}
\caption{MWAX visibility format for the example of four antenna tiles (ant0 to ant3) and two fine channels (ch0 and ch1).  Each visibility value is a complex single-precision floating-point number with real (r) and imaginary (i) components.}
\label{fig:mwax_visibility_format}
\end{center}
\end{figure*}

\subsection{Visibility output ordering} \label{sec:vis_order}

xGPU natively utilises a data ordering that it refers to as \texttt{REGISTER\_TILE\_TRIANGULAR\_ORDER} that is optimised for execution speed.  After an output visibility set has been transferred to the host CPU, it is re-ordered into a more intuitive triangular order referred to as \texttt{MWAX\_ORDER} (slowest-to-fastest changing):  [time][baseline][channel][polarisation].  This format for an individual time-step is illustrated in Figure \ref{fig:mwax_visibility_format} for a simple example of four antenna tiles and two output fine channels.

The reordering to \texttt{MWAX\_ORDER} is performed by the CPU in two stages.  First, a standard xGPU library function is used to convert from \texttt{REGISTER\_TILE\_TRIANGULAR\_ORDER} to xGPU's own \texttt{TRIANGULAR\_ORDER}.  Then a custom function converts from \texttt{TRIANGULAR\_ORDER} to \texttt{MWAX\_ORDER}.  Each visibility set (one per integration time) is written in \texttt{MWAX\_ORDER} to the output ring buffer.

\begin{figure}
\begin{center}
\includegraphics[width=\columnwidth]{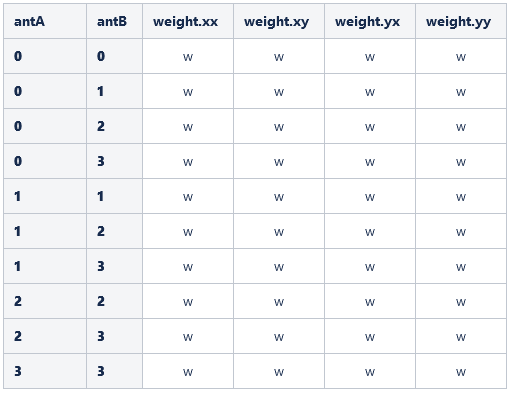}
\caption{MWAX visibility weights format for the example of four antenna tiles (ant0 to ant3).  Each weight value is a single-precision floating-point number that applies commonly to all the fine channels of the specified baseline/polarisation.}
\label{fig:mwax_weight_format}
\end{center}
\end{figure}

Immediately following each visibility set, a table of ``visibility weights'' is written to the output ring buffer, according to the format illustrated in Figure \ref{fig:mwax_weight_format}.  There is a distinct weight value, common across all fine channels, for every baseline/polarisation. Each weight serves as an ``occupancy'' metric, reflecting the fraction of input voltage data present for the signal paths involved in each baseline/polarisation during the integration time for that visibility set.  At the time of writing, these weights are not utilised and all values are set to 1.0.  Their use is categorised as a potential future enhancement and their calculation and intended purpose are explained more fully in \S \ref{sec:vis-weights}.  However, to maintain forward compatibility in correlator output products, dummy weight values are currently written to the output ring buffer and ignored by downstream tools.

\subsection{Data capture, transport, and archiving} \label{sec:archiving}

A separate process running on each MWAX server reads the visibility data from the output PSRDADA ring buffer and writes out a FITS formatted file \citep{1981A&AS...44..363W} to disk. To ensure files do not grow excessively large, the software detects when a file size exceeds a configurable limit (the current setting is 10 GB), and if so will start writing to a new file. 

After each FITS file is written to disk, an MD5 checksum is made from the file and a new entry is made into the MWA M\&C database, recording the file metadata and checksum.  The MD5 checksum type was chosen as it is the native checksum algorithm in use by Pawsey's Ceph-based Long Term Storage (LTS) systems, allowing end-to-end checksum integrity checks using a single method.

Observations designated as a calibrator will then have auto- and cross-correlation statistics produced and sent to the M\&C system as part of ongoing correlator health monitoring. At this point visibilities and voltages are put onto a queue to be sent over a 100 Gbps link via the XRootD protocol\footnote{https://xrootd.slac.stanford.edu/index.html} to the mwacache servers in Curtin's data centre. The ten mwacache servers have combined storage of over 2 PB, allowing data to be buffered in the case where the 100 Gbps link from Curtin to Pawsey is down, or Pawsey itself is unable to ingest the data.

Cached data is transferred to Pawsey's LTS system via the Boto3\footnote{Boto3 provides an API to interact with object stores such as Pawsey's LTS or with Amazon's S3 object store. See: https://boto3.amazonaws.com/v1/documentation/api/latest/index.html} Python library. At any time, eight mwacache servers are used, each receiving and then on-sending data from three of the MWAX Servers. As data are successfully deposited into Pawsey, the system will update the metadata for each file, flagging that it has been ingested successfully.

\subsubsection{MWAX FITS file format}

A standard FITS file with an MWAX specific layout (as opposed to more well known radio astronomy formats) is used for both backwards compatibility with the legacy correlator, and due to the fact that the raw visibilities require some preprocessing before they are science ready. Keeping the file format similar to the legacy correlator simplifies the tooling required to preprocess both legacy and MWAX data, with the preprocessing tools performing of the role of conversion to well-known radio astronomy formats.

The Birli\footnote{https://github.com/MWATelescope/Birli} pipeline is one of the widely used tools (either stand-alone or via the MWA ASVO\footnote{MWA All-Sky Virtual Observatory (MWA ASVO) is MWA's public data portal, allowing users to download and/or preprocess MWA data via a website. https://asvo.mwatelescope.org} data portal) to perform preprocessing to the raw visibilities in both the legacy correlator and MWAX formats. Birli takes all of the FITS files for a coarse channel from each MWAX server, plus a metafits (\S \ref{sssec:observation_parameters}) file, and performs the following operations:
\begin{itemize}
\item{Cable delay corrections;}
\item{Geometric corrections;}
\item{Digital gain corrections;}
\item{Coarse channel passband correction;}
\item{Automated RFI and/or manual flagging of data;}
\item{Calculation of u,v,w coordinates;}
\item{Averaging of visibilities by time or frequency;}
\item{Consolidation of multiple files and coarse channels into a single output dataset;}
\item{Conversion to commonly used, standard radio astronomy formats (currently CASA measurement set \citep{The_CASA_Team_2022} and UVFITS\footnote{Greisen, E. W. 2023, AIPS Memo Series, 117 - AIPS FITS File Format (revised), http://www.aips.nrao.edu/aipsmemo.html} are supported).}
\end{itemize}

Since the preprocessing steps are performed after retrieving the archived data, the MWA community have extensive flexibility and re-processing options with the same raw data, allowing novel usage of the data for different science cases. For example, the same raw data could be averaged by time (for full spectral resolution) by one user or by frequency (for full time resolution) by another.

\subsection{Monitoring and control}

MWAX monitoring and control (M\&C) encompasses all the facilities needed to support the deployment, operation, and monitoring of the MWAX system.

MWAX includes processes running on many separate server nodes: ten Medconv Servers (see \S \ref{sec:medconv}); 26 MWAX Servers (see \S \ref{sec:hardware_networking}) and ten cache boxes in Perth (see \S \ref{sec:archiving}); as well as a few other virtual servers running M\&C tasks that are shared with the rest of the MWA telescope infrastructure.

M\&C for MWAX specifically involves:
\begin{itemize}
\item{Installing, configuring, and maintaining the operating systems on all of the Medconv, MWAX, and cache servers, and compiling/deploying MWAX software to all of the servers after any changes, with the appropriate static configuration settings;}
\item{Starting all the required processes on the Medconv, MWAX, and cache servers as needed, and stopping/restarting en-masse or individually when necessary;}
\item{Passing all of the processes any parameters needed for each observation as it occurs, in real time; and}
\item{Monitoring the health of each process and server to detect any problems.}
\end{itemize}

\subsubsection{Deployment}

All of the MWAX Servers are running Ubuntu 20.04, and because we are heavily input/output bound, we do not use containers to help isolate the correlator processes from operating system changes. To keep all of the servers identical, we run an instance of
``aptly''\footnote{Aptly allows us to mirror and snapshot the official Ubuntu package repositories locally, frozen at a particular point in time. See: https://www.aptly.info}
on site, with a mirror of all Ubuntu 20.04 packages frozen at a particular date. This means that we can reinstall the operating system on a new (or newly repaired) server and still
guarantee that every package will be identical to every other server. If we ever do want to upgrade some or all of the operating system packages, we can create an updated mirror, test it on one of the spare MWAX Servers, then duplicate it on all the other servers with a simple \texttt{apt update} command.

All operating system configuration (creating users, mounting filesystems, etc) and MWAX software deployment (compiling, copying
binaries, configuration files, etc) is done using
Ansible.
With a single command, we can remotely perform everything needed to go from a fresh Ubuntu 20.04 install to a working Medconv, MWAX, or cache server - either on one server at a time, or dozens of servers in parallel.

The same is true for new MWAX software versions - we use one Ansible command to compile and/or deploy new code on a single server for testing, then we repeat the same Ansible command to push it to all of the relevant servers in parallel.

\subsubsection{Startup/shutdown}

All MWAX processes are managed using ``systemd''\footnote{systemd is a popular system and service manager for Linux. See: https://systemd.io}
which handles service startup/shutdown in the correct order, satisfying any dependencies. A single \texttt{mwax\_mandc} process listens for remote commands on a
Redis\footnote{Redis is a popular, open source, in-memory data store used for caching, streaming and data brokerage. See: https://redis.io} message queue and, when commanded, commands systemd on that host to start, stop, or reload any of the MWAX services on that host.

A command line tool on one of the MWA M\&C servers can be used to send the Redis messages needed to start or stop all MWAX processes on all 46 servers in parallel, or start/stop/ reload a single service on a single host, or anything in between.

\subsubsection{Observation parameters (metafits file)}\label{sssec:observation_parameters}

All of the MWAX Servers mount a single shared filesystem via NFS. One directory on that filesystem is reserved for static configuration files, so that they are all in one place. Another directory is reserved for metadata, created in real time by a daemon running on the NFS server host.
That metadata consists of a so-called ``metafits file'' - a single small FITS format file that contains all of the details needed to take (in the case of the correlator) or process (when passed with raw data files) that observation. It is written using the FITS library in Astropy \citep{2013A&A...558A..33A, 2018AJ....156..123A}.

The metadata includes:
\begin{itemize}
\item{The locations of all tiles active in the observation, as well as cable types, cable lengths, etc;}
\item{All per-observation specific data (tile delays, source details, observation name, correlator frequency/time averaging settings, coarse channels used, analog attenuation, etc); and}
\item{Any faults or errors - bad tiles, receivers that failed to communicate when changing channels or attenuation, etc.}
\end{itemize}

The metadata file is created $\approx$4 seconds before each observation starts, with a filename that begins with the observation ID (which is the start time in GPS seconds). It is then available to MWAX processes as data from that observation makes their way through the pipeline. Around 2-3 seconds after the start of each observation, the metafits file is overwritten (atomically, with a filesystem rename operation) with a new copy that contains updated details about any hardware errors that occurred when the observation actually started. Those fault data can then be used for flagging.

The metafits file also contains a table of source Azimuth/ Elevation values calculated every 4 seconds for the duration of the observation, using a software library.
This information can in future be used to calculate the delays needed for each tile to implement fringe stopping.

\subsubsection{Health monitoring}

Every MWAX process, on every server, sends a multicast UDP packet on the 1 Gbps M\&C network, every second. That packet contains health data, ranging from a simple ``I'm alive'' heartbeat message through to JSON structures containing disk free space, buffer availability, etc, depending on the process. A single process on an M\&C server listens for all of the heartbeat packets and maintains a current snapshot of the state of each process, as of the last heartbeat packet received. It makes these status snapshots available via a Redis server, and they are used to generate dynamic web pages showing MWAX's status and health, as well as other diagnostic tools.

The overall server and environmental status monitoring uses Icinga\footnote{https://icinga.com}, an open-source tool that we have configured to monitor rack temperature, server power use and temperatures, disk free space, network time synchronisation status, disk errors, load average, etc. If there are any problems, the MWA Operations team is notified by email, and when necessary, Icinga will take action itself (shutting down an entire rack of servers if it gets too hot, for example).

\subsection{Offline correlation}

MWAX has been designed such that the voltage subfiles can be replayed offline for the purposes of offline correlation or testing, whether on the dedicated MWAX Servers at the MRO or other GPU-equipped servers.  This is made possible because the same modular code used for real-time operation on the MWAX Servers can also be installed on other compute nodes with different specification GPUs (e.g. lower specification GPUs at Pawsey) to provide an offline mode that will typically operate below real-time speed. Note that the time/frequency averaging modes available when running MWAX offline depend heavily on the server and GPU hardware it is executed on.

Subfiles correlated offline can be reprocessed with different parameters to those requested for the original observation, for example with different time or frequency averaging. Subfiles are constructed with a margin buffer of voltages extending 4096 samples beyond the start and end of the sub-observation.  This allows observations to be reprocessed with different whole-sample delays, or existing delays to be removed using external software.

\section{MWAX VALIDATION AND BENCHMARKING}
\label{sec:validation}

\begin{figure*}[t]
\begin{center}
\includegraphics[width=465pt]{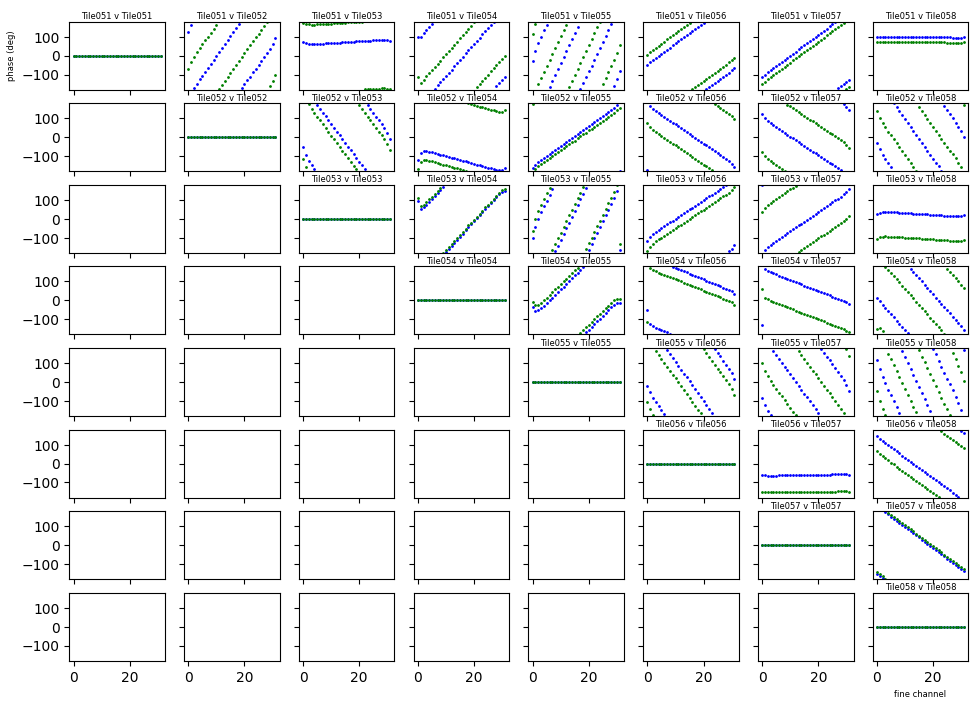}
\caption{Phase versus frequency plots for 36 baselines demonstrating clear fringes during a Sun test observation.  The two colours represent two different polarisation cases: correlation of x versus x, and correlation of y versus y.
}
\label{fig:phase_plot_mwax}
\end{center}
\end{figure*}

\subsection{Hand-crafted test vectors}

A number of test data sets were artificially constructed and injected into MWAX as test vectors.  These data consisted of apparent pure tones of known frequencies, amplitudes and phases for each input signal path.  Code independent of MWAX was used to compute the expected visibility output once correlated.  The MWAX output was compared with those results to confirm the underlying mathematical processing and that the visibility output orderings were consistent with each other.

The test data sets (encompassing one coarse channel, for one 8 second sub-observation, for 128 dual polarisation inputs) were constructed using a combination of a pseudo random number generator and a simple digital phase rotator.  For each polarisation of each tile, a starting complex ``voltage'' time sample was generated.  Subsequent time samples for that polarisation were generated by applying a fixed rotation to each sample, digitally simulating a pure frequency tone at a fine channel that was selectable by altering the rotation angle used. These data were placed into a standard subfile format and passed into the MWAX processing pipeline for fine channelisation, cross-correlation and time integration.  The resultant MWAX visibilities were then compared against the independently computed visibilities.

We had a high degree of confidence in the `tried and tested' NVIDIA cuFFT and xGPU libraries. Our primary goal with these tests was to verify that the data passed into these libraries were correctly ordered and free from such possible bugs as `off by one' errors, and to ensure the visibility output ordering after time and frequency integration matched expectations. A secondary goal was to compare the numerical accuracy of MWAX’s single-precision floating point calculations against independent double-precision results.

In all trials, the visibilities showed the expected time integrated results, at the correct fine channel, and for the correct baseline with no unexpected digital cross-contamination between signals as might be expected if data sequence or data ordering bugs were present.  The notable exception to this was the imaginary components of parallel-polarisation autocorrelation quantities (i.e., the XX and YY autocorrelations of tile inputs), where very small but non-zero residual values were present.  These were initially predicted to be exactly zero and while the values seen were too small to affect science operation, their cause was investigated for completeness. Communication with NVIDIA’s xGPU technical team provided an explanation of how this is the result of subtle floating point rounding within the CUDA single-precision fused-multiply-add operation.

\subsection{Visibility spectra and fringes}

Early validation of the MWAX correlator involved plotting phase versus frequency of raw visibilities for a subset of 36 baselines for a single integration and coarse channel of an observation where the MWA was pointed at the Sun. The visibilities were read from their native FITS formatted data file using a python script and were not pre-processed or corrected in any way for cable lengths nor antenna layout/geometry. The plots produced (see Figure \ref{fig:phase_plot_mwax}) show the expected flat phase versus frequency centred on zero degrees for the autocorrelations and also very obvious fringes for the cross-correlations.

\begin{figure}
\begin{center}
\includegraphics[width=200pt]{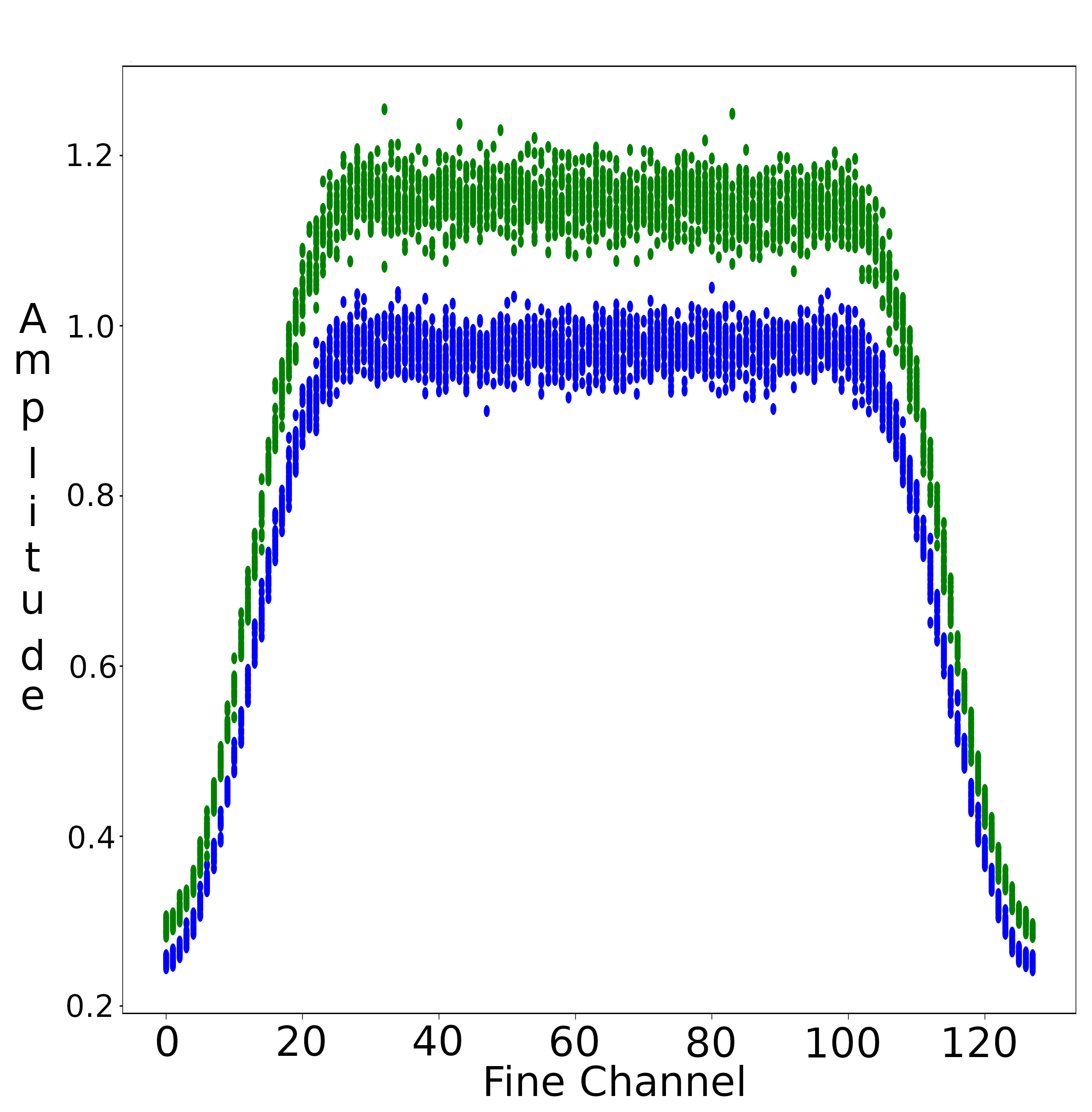}
\caption{Amplitude versus fine channel plot for a single coarse channel and cross-correlation baseline for 10 seconds integration time, demonstrating the expected MWA receiver band-shape for a Hydra A test observation. The two colours represent two different polarisation cases: correlation of x versus x, and correlation of y versus y.}
\label{fig:ppd_plot_mwax}
\end{center}
\end{figure}

Another early validation method was to verify that the typical MWA coarse channel band-shape (high attenuation in the first 10\% and last 10\% of fine channels in each coarse channel) was present in the raw visibilities (again, no corrections were applied). As can be seen in Figure \ref{fig:ppd_plot_mwax} in this amplitude versus fine channel plot for a single cross-correlation baseline, twenty 0.5 second integrations (10 seconds total), and coarse channel, the MWA receiver band-shape is clearly present.

This early, simplistic validation work showed that the MWAX visibilities were seeing real data from the telescope, that the clocking system was operating correctly to allow for correlation of signals from different receivers, and that the visibilities were in the expected format and order.  Having established this, more comprehensive validation work would follow.

\subsection{Cross-validation against the legacy correlator}

The MWA's legacy correlator was in operation from 2013 until early 2022 and was a trusted instrument, albeit with some known shortcomings.
This included a fine channel passband shape that did not result in equally weighted contributions to visibilities from all frequencies in the channel (see \S \ref{sec:averaging}). There was also degraded accuracy as a result of the various quantisation stages and asymmetric rounding within its FPGA fine channeliser PFB \citep{Mcsweeney2020}.
Due to both of these effects, it was fully expected that MWAX would not produce visibilities that exactly matched the legacy system, but should in principle provide improved accuracy. 

To facilitate direct comparison between the legacy and MWAX correlators, an optical splitter was used to duplicate one of the three fibres coming from each receiver. The duplicated optical signal was connected directly to an unused input port on the EDT cards while the system was still in the legacy configuration. This allowed a copy of the receiver coarse channel data to bypass the fine PFB and be captured and re-packetised as a prototype for the full MWAX system. These duplicated data contained one third of the bandwidth of the full system.
The duplicated, re-packetised data were streamed simultaneously to a test system running MWAX while the legacy correlator was running, providing a mechanism to compare the outputs on identical live data.

The most direct comparison between systems was to correlate with 10\,kHz frequency resolution, where the output of the legacy fine PFB is unaltered.
Although the visibilities were not expected to be identical, they could be expected to be within a few percent of each other.

\begin{figure}[t]
\begin{center}
\includegraphics[width=250pt]{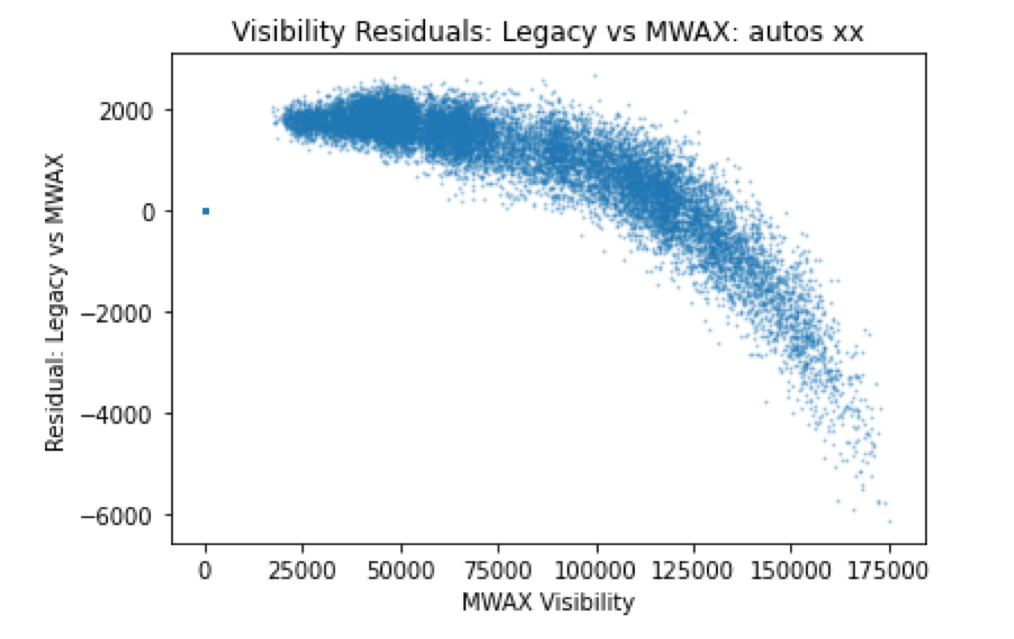}
\caption{Example of the differences between legacy and MWAX visibilities as a function of the MWAX visibility.  This plot is for all the x-polarisation autocorrelations (real component).}
\label{fig:residuals_autos}
\end{center}
\end{figure}
\begin{figure}
\begin{center}
\includegraphics[width=255pt]{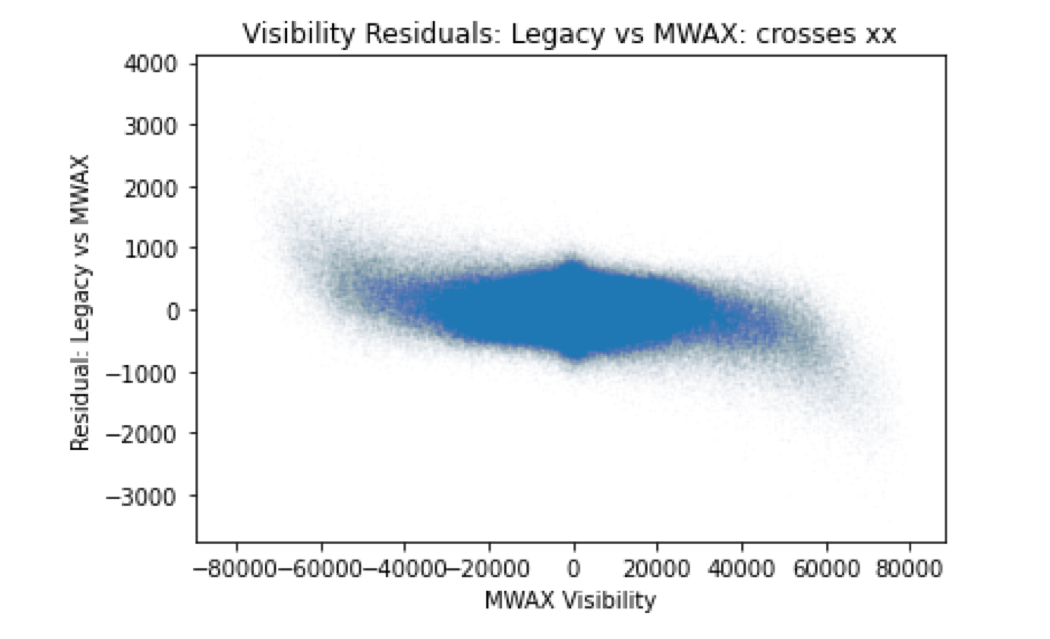}
\caption{Example of the differences between legacy and MWAX visibilities as a function of the MWAX visibility.  This plot is for all the x-polarisation cross-correlations (real component).}
\label{fig:residuals_crosses}
\end{center}
\end{figure}

For several different observations, the following were conducted:

\begin{itemize}
\item{Magnitude and phase plots as a function of fine channel were visually compared to confirm similarity of shape and the slope of visibilities;}
\item{Scatter plots were formed of MWAX visibilites versus legacy visibilities for autocorrelations and cross-correlations (for both the real and imaginary components).  Perfect agreement would result in all comparison points falling on a straight line of slope 1 (following magnitude normalisation).  The actual plots showed straight lines of slope 1, slightly broadened (as expected due to value differences) and no outlier points; and}
\item{The differences between normalised visibility values (``residuals'') were plotted as a function of the MWAX visibility values. (MWAX was used as the reference because it was assumed to be the more accurate).  Perfect agreement would result in a horizontal line at zero residual.  Examples of the actual plots are given in Figure \ref{fig:residuals_autos} (for autocorrelations) and Figure \ref{fig:residuals_crosses} (for cross-correlations).  Again, there were no unexpected outlier points, and the deviations from ideal can all be explained by the abovementioned passband/quantisation/rounding effects, as we do in detail below.}
\end{itemize}

With reference to Figure \ref{fig:residuals_autos}, the main features are a large positive offset (even at low visibility magnitudes), a pronounced ``drooping'' with increasing visibility magnitude, and a broadening of the ``line''.  The offset is a result of the internal quantisations and rounding within the legacy fine channeliser PFB (the squaring operation for autocorrelations means that quantisation/rounding errors will accumulate).  The drooping is a result of saturation/clipping effects, again taking place within the legacy fine channeliser PFB and not observed with MWAX since it uses 32-bit floating-point arithmetic within its F-engine.  The legacy values are smaller than they should be with large visibility magnitudes, resulting in the observed downswing of residuals.  The line broadening can be explained as a combination of quantisation effects and differences between the fine channel passband shapes of the two F-engines.

With reference to Figure \ref{fig:residuals_crosses}, the main features are the symmetry of residuals around zero, drooping at larger visibility magnitudes (upwards for negative visibilities and downwards for positive visibilities), and a broadening of the ``line''.  The symmetry is expected because, for cross-correlations, the quantisation and rounding errors average away. The drooping is again due to saturation/clipping effects within the legacy fine channeliser PFB.  The legacy values are less negative than they should be with large negative visibility magnitudes, resulting in an upswing of residuals on negative visibilities, and less positive than they should be with large positive visibility magnitudes, resulting in a downswing of residuals on positive visibilities.  The line broadening is for the same reasons as explained for autocorrelations, and the width of the broadening is seen to be similar in both cases.

In summary, the observed residuals contained no outliers and could all be explained by design differences between the two correlator systems, giving confidence that MWAX's performance is consistent and accurate.

\subsection{Performance benchmarking}
\label{sec:benchmarking}

A key requirement for MWAX is sustained real-time operation with up to 256 antenna tiles.  The configuration of the MWAX Server hardware described in \S \ref{sec:hardware_networking} was scoped accordingly, based on benchmarking tests conducted using test subfiles for 256 tiles (512 signal paths).  These subfiles needed to be `manufactured' by duplicating signal paths, as the MWA itself was receiver-limited to just 128 tiles.

\begin{table*}[!t]
\begin{center}
\begin{tabular}{l c c c}
\hline
                                              & & Integration Time / Fine Channel Bandwidth \\
\hline
Sub-function                                  & 250 ms / 40 kHz & 1 s / 10 kHz & 8 s / 1 kHz \\
\hline \hline
cuFFT fine channelisation                     & 0.5 s & 0.5 s & 0.5 s \\
Delay and phase corrections                   & 1.3 s & 1.3 s & 1.3 s \\
Time/frequency sample reordering              & 0.4 s & 0.4 s & 0.4 s \\
xGPU cross-correlation                        & 2.1 s & 2.1 s & 2.1 s \\
Channel averaging                             & 0.1 s & 0.1 s & 0.1 s \\
Visibility reordering to MWAX format          & 0.6 s & 0.7 s & 1.8 s \\
Bus traffic and miscellaneous                 & 1.2 s & 0.8 s & 1.3 s \\
\hline
Total execution time                          & 6.2 s & 5.9 s & 7.5 s \\
\hline
\end{tabular}
\caption{Example FX-engine execution times for a single 256-tile subfile, running on the MWAX Server hardware specified in \S \ref{sec:hardware_networking}.}
\label{table:benchmarks}
\end{center}
\end{table*}

Table \ref{table:benchmarks} presents execution times for the MWAX FX-engine, broken down by key sub-function, for three example correlation mode settings.  Fractional delay and phase corrections have been included, although these features are not officially supported at the time of writing.  These are the times required to process a single 8 second subfile, so real-time operation is maintained when the total execution time is below 8 seconds.

\subsection{MWAX commissioning and first light}
\label{sec:commissioning}

The main goal of Engineering Commissioning was to test the correctness and stability of MWAX across a range of observing modes. In addition to the direct visibility comparisons described above, a matrix of observations was generated covering combinations of observing frequency, frequency resolution, time resolution, and length of observation. Sources were chosen for maximum diagnostic value, including the Sun, Galactic centre, strong compact ``A-team'' sources, and well-known geosynchronous satellites that have well defined frequency structure in the 250--280\,MHz frequency range.

The visibilities from these test observations were manually inspected as raw data, and also after being processed by the MWA's data conversion pipelines \citep{2015PASA...32....8O}, to ensure that metadata, signal mappings, and corrections (to create a phase centre) were applied correctly.
This process flushed out some minor bugs and found only one correctness issue: a sign convention problem on the XY autocorrelation products which has been in the system for some time, but never noticed because the XY autocorrelations are rarely, if ever, used.

\section{ROADMAP FOR FUTURE DEVELOPMENTS}
\label{sec:future}

\subsection{\label{sec:vis-weights} Visibility weights}

Any UDP-based communications network can experience occasional packet losses due to network congestion, buffer overruns, link errors etc. A potential future addition to the MWA is pre-correlation RFI flagging, where certain packets may be flagged and excised.
In both cases, the UDP Capture code on each MWAX Server may construct subfiles that contain missing data blocks, where the missing data have been replaced with zero-valued samples.
In such cases, the UDP Capture code will set corresponding ``zeroing flags'' within a metadata table that is placed in block 0 of the subfile.

Each visibility set generated by the correlator has an associated array of ``visibility weights'' that are placed in the output ring buffer, interleaved with the visibility values themselves (see \S \ref{sec:vis_order}).  There is one floating-point weight associated with each baseline/polarisation, which applies to all fine channels for that baseline/polarisation.  This value is intended to range from 0.0 to 1.0, representing the fractional data occupancy, i.e. the fraction of the integration time for that visibility set that contained fully populated 50 ms sub-blocks in both signal paths for that baseline/polarisation.  The visibility values can optionally have their magnitude normalised using these weights such that missing packets do not cause variations in visibility magnitudes. Since the weights that have been applied for this normalisation process are written to the output ring buffer, downstream tools are able to access the values to understand exactly how the corresponding visibilities were computed, which may be desirable for some applications, e.g. where signal-to-noise ratios of visibilities are meaningful. At the time of writing, all output visibility weight values are set to 1.0.  Current packet loss rates are well below 0.1\% so this normalisation stage is not consequential.  However, to allow for possible future increases in packet loss rate (e.g. when moving to a higher number of correlated tiles) and/or RFI excision, it is intended to add a feature where the correlator tracks the zeroing flags of incoming subfiles and computes dynamically the corresponding weight to be applied to each visibility.

\subsection{Coarse channeliser PFB de-ripple}

The critically-sampled coarse channeliser PFB of the existing receivers employs short (8-tap) constituent filters and hence the passband frequency response requires relatively wide transition bands to manage spectral leakage, and exhibits a degree of passband ripple \citep{Prabu2015}.  This passband response is deterministic and can be compensated for during downstream signal processing to provide a flatter response over a greater portion of the coarse channel band.
This may be of value to science cases that make power spectra (or similar) from visibilities, where calibration is not performed independently across fine channels.

One logical place to accomplish this ``de-ripple'' or ``equalisation'' is within the correlator.  The MWAX F-engine includes a fractional delay correction stage where every spectral sample of every input signal path is multiplied by a complex gain taken from a pre-computed look-up table.  Presently the table contains linear phase gradients corresponding to different fractional delay values.  These gradients could be combined with a spectral magnitude response that represents the inverse of the receiver PFB's passband response.  As the delays are applied, the spectral samples will then simultaneously be magnitude-adjusted to provide a flattened overall frequency response.

This correction assumes that the telescope sensitivity is sufficiently constant across a coarse channel, and the corrections are relatively small, such that there is no impact on visibility weights.

For future receivers that employ oversampling (see \S \ref{sec:oversampling}) and/or a different coarse channel passband response, the delay look-up table contents would be updated accordingly.  Heterogeneous sets of receiver types could be accommodated by having separate look-up tables for each passband response employed (although this would require extending the metadata passed to the correlator to inform it which receiver type corresponds to each input signal path, which is not currently supported).

\subsection{Cable delay correction and fringe stopping} \label{sec:delay-correction}

As already mentioned in \S \ref{sec:int_delay} and \S \ref{sec:f_engine}, MWAX has been designed to support delay corrections to each signal path prior their cross-correlation, such that each signal path can be ``phased up'' to a specific direction on the sky known as the ``correlation pointing centre''. The (coarse channel) whole-sample delay components can be corrected for within the UDP Capture stage, and residual fractional sample delay components can be corrected for within the F-engine.

The delays introduced by varying cable/fibre lengths for data and clock signals are relatively stable for a given array configuration, aside from minor drifting due to age and temperature changes.  These delay values are stored in the metafits file of each observation and are regarded as static, with the minor variations being compensated for in the post-correlation calibration stage.

When pointing the telescope at a fixed RA/Dec on the sky, the Earth's rotation results in dynamically changing geometric delays between sky sources and each tile.  Without any corrections for this effect, individual visibility values will drift in phase, as will the ``fringe'' plots of phase versus frequency.  Correcting for this effect to maintain constant phases as the Earth rotates is referred to as ``fringe stopping''. 

For the MWA extended array's $\sim 5$\,km baselines at 300\,MHz ($\lambda=1$\,m), the fringe frequency is $\sim 2$ radian/sec \citep[][their eq. 4.9]{2017isra.book.....T}.  Without fringe stopping there will always be some degree of decorrelation during a time integration interval.  For this reason the maximum time averaging used in the legacy correlator for extended array observations was chosen as 0.5\,sec, which limited the worst-case decorrelation (experienced on the longest baselines at 300 MHz) to 5.4 \% \citep{Wayth2018}.  Decorrelation can be reduced to effectively zero (at the phase centre) on all baselines once fringe stopping is introduced.

Both static corrections for cable delays and dynamic corrections for fringe stopping will be able to be handled through the common mechanism of whole/fractional delay correction described in \S \ref{sec:int_delay} and \S \ref{sec:f_engine}, which supports distinct delay values for every signal path at every 5 ms timestep, i.e. 1600 values over an 8 second sub-observation.  In the case of fringe stopping, every one of the 1600 values is likely to be different.  If fringe stopping is not enabled and only static cable delay correction is performed, each of the 1600 values will be the same.

The whole and fractional delays for each signal path will be recalculated independently for each sub-observation across the entire observation.

At the time of writing, tests of the static cable delay corrections had been successfully completed by taking test observations on calibration sources as a voltage capture, then offline processing the data with and without the static delay corrections in place. The downstream data conversion tool, Birli, can optionally apply the corrections post correlation as a frequency-dependent phase rotation, the resulting visibilities having flat phase as a function of frequency. The tests showed virtually identical results between MWAX-applied static delay corrections and the Birli-corrected visibilities.

Full support for delay correction will be provided in a future MWAX software release.  At that time, cable delay correction and fringe stopping will be made available as optional user settings when scheduling observations.


\subsection{Real-time tied-array beamforming}

Tied-array beamforming is a method of spatial filtering where-by the field-of-view of an array telescope is narrowed around a particular pointing direction whilst increasing the sensitivity in that direction.  It is accomplished by coherently summing the voltage signals from multiple antennas, first delay-adjusting each signal path such that all are phase-aligned for a signal originating from the beam's boresight. It is commonly used for performing high-time-resolution analysis of signals from specific sources on the sky, such as in pulsar science.  In the legacy MWA system it was necessary to perform beamforming through the offline processing of VCS data \citep{tremblay15}. This approach necessitates the storage of large volumes of voltage data. An attractive alternative is to perform the beamforming online in real-time, which affords a major reduction in data storage since each output beam has a data volume equivalent to only a single dual-polarisation antenna.

As discussed in \S \ref{sec:delay-correction}, the fractional delays required for all signal paths to be phased to a given correlation pointing centre are passed to the correlator in the metadata block of the subfile.  In a similar way, additional delay metadata may be placed in the metadata block corresponding to other pointing directions.  In this way, the same subfile may serve as input to a tied-array beamforming function where delay-corrected signal paths are coherently summed rather than correlated.

A GPU-accelerated real-time multibeam beamformer of this type is currently in development.  It will simultaneously form, from the same subfile, multiple incoherent and coherent beams.  The incoherent beams will all have the same field-of-view (the full field-of-view of the tiles' analog beamformers), but will provide differing frequency/time resolutions to suit differing applications.  The coherent beams will all be at the full bandwidth and time resolution of a coarse channel, but with differing pointing directions.

Integration of the multibeam beamformer into MWAX Servers will be straighforward because both the correlator and beamformer processes can attach to a common input ring buffer that is ingesting the subfiles generated during an observation.

\subsection{MWA Phase III and support for oversampled coarse channels}
\label{sec:oversampling}

A known limitation of the existing MWA signal chain is the use of critically-sampled PFBs for coarse channelisation within the current receiver systems.  The passband response of coarse channels exhibits steep attenuation at the channel edges, which results in approximately 10\% of fine channels being effectively unusable.  Across a contiguous band of coarse channels, not all frequencies contribute equally to visibility outputs, and any attempt to reconstruct higher time-resolution signals by aggregating coarse channels will inevitably be compromised.

As part of the MWA Phase III development programme, new receiver designs are being developed that will have the capability of overcoming this deficiency.  Optionally, coarse channelisation will be able to be performed using an oversampled PFB, which will allow a flat frequency response to be achieved across the entire coarse channel passband (equal to the channel spacing).  This will ensure no fine channels across the passband are lost, and ultra-high time-resolution will be possible through high-fidelity PFB inversion, which can be accomplished more readily with oversampling \citep{Morrison2020, Bunton2021}.

This extension to using oversampled coarse channels will have impacts on MWAX in several areas, including UDP Capture and the construction of subfiles, as well as requiring a more sophisticated F-engine that can properly accommodate oversampling.  MWAX uses FFTs in the F-engine to generate ultra-fine channels that are subsequently aggregated to form fine channels.  This approach facilitates a straightforward method of dealing with oversampling, since those ultra-fine channels constituting the oversampled portions of each coarse channel can simply be discarded before passing spectral domain data to the X-engine.

\subsection{SKA-Low / MWA bridging}

The MWA is a precursor instrument to the low-frequency component of the Square Kilometre Array (SKA-Low).  Its sensitivity and beam profiles are well understood and characterised, allowing effective calibration and accurate measurement of absolute source flux densities.

At the time of writing there are two operational SKA-Low prototype arrays: AAVS2 and EDA2 \citep{2020SPIE11445E..89V,2022JATIS...8a1014M,2022JATIS...8a1010W}, which are co-located with the MWA at the MRO.  Each array is the equivalent of one SKA-Low station, consisting of 256 dual-polarisation antennas, with a common suite of analogue and digital downstream hardware.

SKA-Low verification activities can benefit from cross-correlation of these prototype stations with MWA tiles. Historically, the MWA has provided the primary mechanism to verify the function and sensitivity of early SKA-Low pre-prototype systems AAVS0.5 and EDA1 \citep{2015ITAP...63.5433S,2017PASA...34...34W}.
For AAVS0.5 and EDA1, the analogue signals from the prototype systems were attached to an MWA receiver, replacing the analogue signal from an MWA tile.
The MWAX architecture allows packetised data to be directly injected into the main Nexus 9504 switch as shown in Fig. \ref{fig:MWAX_signal_path}, which will allow the MWA to accept and cross-correlate data from additional sources without the need to replace the analogue signal from a tile.

In the case of SKA-Low, however, the sampling rates and coarse channel bandwidths of SKA-Low (800 Msamp/s, 0.926\,MHz) and the MWA (655.36 Msamp/s, 1.28\,MHz) are different, preventing direct cross-correlation.  A bridging system is in development that will re-sample the SKA station beam signals to match the MWA specification, including packetising the SKA data into the MWA's Ethernet packet format. 
These additional signal paths can be added to the MWA metadata and all MWAX Servers will recognise them and generate subfiles containing the additional signal paths - which the correlator will then automatically process as if they were additional MWA signal paths.

The bridging method involves ultrafine channelisation of the oversampled SKA coarse channels, then selection of subsets of these ultrafine channels to form a contiguous set corresponding to MWA coarse channels, then inversion to form the coarse channel. This process is non-trivial because the coarse channel centre frequencies do not align between SKA-Low and MWA coarse channels.

Aside from assisting with SKA-Low verification activities, combining these two SKA stations with MWA tiles will provide a substantial sensitivity improvement to the MWA, raising its total collecting area by $\approx25\%$ over its 128 tile configurations.

\subsection{Expansion to SKA-Low scale correlation}
\label{sec:SKA}

Whilst there is no intention of deriving a correlator solution for SKA-Low from MWAX, it is instructive to consider what would be required if that were ever contemplated.

MWAX was scoped and benchmarked for real-time correlation of up to 256 dual-polarisation tiles.  SKA-Low is anticipated to consist of 512 dual-polarisation stations.  The cross-correlation of station beams will therefore require a correlator that accepts twice the number of input signal paths as MWAX. However, SKA-Low coarse channels have a smaller sample rate of $\approx$0.93 Msample/s as compared to 1.28 Msample/s for the MWA.  This implies the input data ingest rate is only increased by $\approx$45\%, and there would be a similar increase in the computational workload for the F-engine (per coarse channel).  Additionally there would be an $\approx$2.5x increase in the workload for the X-engine, due to a four-fold increase in the number of baselines and a smaller channel bandwidth of $\approx$781 kHz.

Taking account of the respective shares of the overall workload in the F- and X-engines, we estimate approximately an overall 2.2x increase in GPU workload as compared to MWAX per coarse channel.  Furthermore, taking the specific example of 10 kHz fine channels and an integration time of 1 second, MWAX benchmarking found that 8 seconds of data from 256 tiles could be correlated faster than real-time, in $\approx6$ seconds.  If we have all 8 seconds available, the implication is that this same mode with SKA-Low could be supported with a GPU of $\approx$1.7x the capability of the A40 GPUs of MWAX.  A doubling of GPU capacity should be sufficient to support a wide range of operating modes.  The number of compute nodes for SKA-Low would be larger, scaled to however many coarse channels are to be instantaneously correlated.

MWAX voltage capture relies on the parallel write performance from twelve hard disk drives in RAID 5 and can achieve up to 2.0 GB per second sustained, sequential write throughput per server/coarse channel. The $\approx$45\% increase in ingest data rate increases the disk write speed requirements for voltage capture operations beyond what this disk system can achieve. For the SKA-Low case, upgrading to using existing solid-state drive (SSD) or non-volatile memory express (NVMe) technology should comfortably scale to, and achieve up to and over 3.0 GB per second (although for the capacities required to allow on-site buffering, the solution will be considerably more expensive per terabyte). Even with the expected increase in visibility data rates out of the correlator, writing visibilities to disk is less demanding than voltage capture, and the existing MWAX disk system would likely be sufficient for this task for most use cases. 

To summarise, the primary extensions required for MWAX to support SKA-Low station correlation would be: (1) a $\approx$45\% increase in the input data ingest rate, (2) roughly a doubling of GPU compute node capability, and (3) an NVMe or SSD based disk system for voltage capture.  The full SKA-Low correlator will not be required for several more years - two or three generations of GPU, disk and server development into the future - so these requirements do not seem onerous.

\section{SUMMARY}
\label{sec:summary}

MWAX is a versatile, scalable, and highly maintainable correlator solution for the MWA.  It replaced the legacy MWA correlator in early 2022 after successful validation and commissioning campaigns.

Due to limited access to receiver systems, the initial deployment correlated only 128 antenna tiles, as with the legacy correlator.  Two additional receivers are in the process of being commissioned, which will extend the MWA to 144 tiles.  By design MWAX has the flexibility and compute capacity to correlate 256 tiles (and more in certain modes).  Extension to 256 tiles and beyond will be rolled out as additional new receivers become available.

Beyond supporting more tiles, MWAX provides a number of additional benefits over the legacy correlator, including higher accuracy (by using floating-point arithmetic to reduce quantisation/saturation degradations), and greater flexibility in the choice of time and frequency resolutions.

A roadmap of potential future functional developments has also been presented, including weighting of output visibilities to compensate for lost or excised UDP packets, coarse channeliser PFB de-ripple, cable delay correction, and fringe stopping.  Integration of coherent tied-array beamforming with the correlator is also in development, as is preparation for accommodating oversampled coarse channels from future receivers.  Work is also underway on an SKA-Low/MWA Bridge system that will allow SKA-Low station beams to be correlated with the MWA, which will be a valuable tool in support of future SKA-Low validation and calibration activities.

Finally, we note the potential for the MWAX design to be scaled to support other demanding correlation applications, such as future SKA-Low station correlation.

\section*{Acknowledgements}
The authors wish to acknowledge valuable early design contributions made by Andrew Jameson.  They also thank the Critical Design Review team led by Adam Deller, and are grateful for many improvements to the manuscript suggested by Danny Price and the anonymous reviewer.

Support for the purchase of the compute hardware was provided via a research infrastructure upgrade grant under the auspices of the National Collaborative Research Infrastructure Scheme (NCRIS) provided by the Australian Government, administered by Astronomy Australia Limited (AAL). 

This work uses data obtained from Inyarrimanha Ilgari Bundara / the Murchison Radio-astronomy Observatory. We acknowledge the Wajarri Yamaji People as the Traditional Owners and native title holders of the Observatory site. Establishment of CSIRO's Murchison Radio-astronomy Observatory is an initiative of the Australian Government, with support from the Government of Western Australia and the Science and Industry Endowment Fund. Support for the operation of the MWA is provided by the Australian Government (NCRIS), under a contract to Curtin University administered by AAL. This work was supported by resources provided by the Pawsey Supercomputing Research Centre with funding from the Australian Government and the Government of Western Australia.

\bibliography{mendeley_references, other_mwax_references}


\appendix

\section{Available Correlation Modes for 128 Tiles}

\begin{figure*}[t]
\begin{center}
\includegraphics[width=470pt]{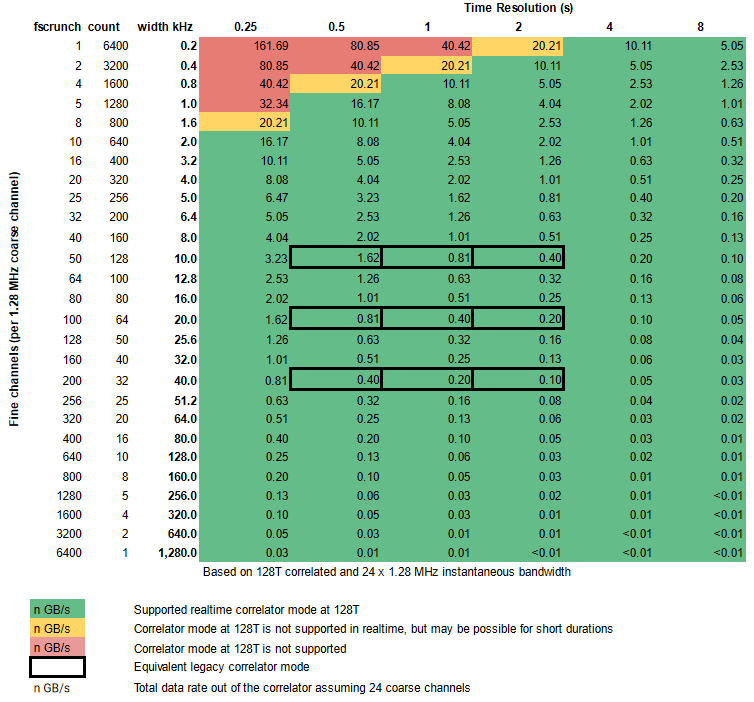}
\caption{Matrix of supported correlator modes for a 128 tile MWA configuration.}
\label{fig:mwax_modes_128T}
\end{center}
\end{figure*}

Figure \ref{fig:mwax_modes_128T} tabulates the frequency and time resolution options available when operating MWAX with 128 tiles.

\end{document}